\documentstyle[twoside,psfig,ijmpa1]{article}

\catcode`\@=11
\long\def\@makefntext#1{
\protect\noindent \hbox to 3.2pt {\hskip-.9pt  $^{{\eightrm\@thefnmark}}$\hfil}#1\hfill}
\def\@makefnmark{\hbox to 0pt{$^{\@thefnmark}$\hss}}	
\def\ps@myheadings{\let\@mkboth\@gobbletwo
\def\@oddhead{\hbox{}
\rightmark\hfil\eightrm\thepage}   
\def\@oddfoot{}\def\@evenhead{\eightrm\thepage\hfil
\leftmark\hbox{}}\def\@evenfoot{}
\def\sectionmark##1{}\def\subsectionmark##1{}}

\def\qed{\hbox{${\vcenter{\vbox{			%HOLLOW SQUARE
   \hrule height 0.4pt\hbox{\vrule width 0.4pt height 6pt
   \kern5pt\vrule width 0.4pt}\hrule height 0.4pt}}}$}}

\renewcommand{\thefootnote}{\fnsymbol{footnote}}  %USE SYMBOLIC FOOTNOTE

\newcommand{\be}{\begin{equation}}
\newcommand{\bea}{\begin{eqnarray}}
\newcommand{\ba}{\begin{array}}
\newcommand{\ea}{\end{array}}
\newcommand{\ee}{\end{equation}}
\newcommand{\eea}{\end{eqnarray}}

\begin{document}
\setlength{\textheight}{7.7truein}  %for 2nd page onwards

\thispagestyle{empty}

%\markboth{\protect{\footnotesize\it Instructions for Typesetting
%Manuscripts}}{\protect{\footnotesize\it Instructions for
%Typesetting Manuscripts}}

\normalsize\textlineskip

\setcounter{page}{1}

\copyrightheading{}		%{Vol.~0, No.~0 (2000) 000--000}

\vspace*{0.88truein}

%\fpage{1}

\centerline{\bf CASIMIR EFFECTS IN RENORMALIZABLE  }
\vspace*{0.035truein}
\centerline{\bf QUANTUM FIELD THEORIES\footnote{Based on talks presented
by the authors at the $5^{\rm th}$ workshop `QFTEX', Leipzig, Sept.~2001.}}
\vspace*{0.035truein}
\vspace*{0.37truein}
\centerline{\footnotesize NOAH GRAHAM}
\baselineskip=12pt
\centerline{\footnotesize\it Department of Physics and Astronomy,
University of California at Los Angeles}
\baselineskip=10pt
\centerline{\footnotesize\it Los Angeles, CA 90025}

\vspace*{10pt}
\centerline{\footnotesize ROBERT L. JAFFE}
\baselineskip=12pt
\centerline{\footnotesize\it Center for Theoretical Physics, Laboratory for
Nuclear Science, and Department of Physics}
\baselineskip=10pt
\centerline{\footnotesize\it
Massachusetts Institute of Technology}
\baselineskip=10pt
\centerline{\footnotesize\it Cambridge, MA 02139}

\vspace*{10pt}
\centerline{\footnotesize HERBERT WEIGEL\footnote{Heisenberg--Fellow}}
\baselineskip=12pt
\centerline{\footnotesize\it Institute for Theoretical Physics,
T\"ubingen University} \baselineskip=10pt
\centerline{\footnotesize\it Auf der Morgenstelle 14 D--72076
T\"ubingen, Germany}

\vspace*{0.225truein}
%\publisher{(received date)}{(revised date)}

\renewcommand{\thefootnote}{\alph{footnote}}

\abstracts{ We review the framework we and our collaborators have
developed for the study of one--loop quantum corrections to extended
field configurations in renormalizable quantum field theories.  We
work in the continuum, transforming the standard Casimir sum over
modes into a sum over bound states and an integral over scattering
states weighted by the density of states.  We express the density of
states in terms of phase shifts, allowing us to extract divergences by
identifying Born approximations to the phase shifts with low order
Feynman diagrams.  Once isolated in Feynman diagrams, the divergences
are canceled against standard counterterms.  Thus regulated, the
Casimir sum is highly convergent and amenable to numerical
computation.  Our methods have numerous applications to the theory of
solitons, membranes, and quantum field theories in strong external
fields or subject to boundary conditions.  }{}{}

\baselineskip=13pt

\section{Introduction}

In these talks, we describe our project to develop reliable, accurate,
and efficient techniques for a variety of calculations in
renormalizable quantum field theories in the presence of background
fields.  These background field configurations need not be solutions
of the classical equations of motion.  Our calculations are exact to
one loop, allowing us to proceed where perturbation theory or the
derivative expansion would not be valid.  For example, in a model with
no classical soliton we can demonstrate the existence of a
non-topological soliton stabilized at one loop order by quantum
fluctuations.  We renormalize divergences in the conventional way: by
combining counterterms with low order Feynman diagrams and satisfying
renormalization conditions in a fixed scheme.  In this way we are
certain that the theory is being held fixed as the background field is
varied.  Our methods are also efficient and practical for numerical
computation: the quantities entering the numerical calculation are
cutoff independent and do not involve differences of large numbers. 
The numerical calculations themselves are highly convergent.

Our methods are limited to one loop and, except in special cases, to
static field configurations.  We also require the background field
configuration to have enough symmetry that the associated scattering
problem admits a partial wave expansion.  The one--loop approximation
includes all quantum effects at order $\hbar$.  It is a good
approximation for strong external fields or when the number of
particles circulating in the loop becomes large.  Even when it cannot
be rigorously justified, the one--loop approximation can provide
insight into novel structures in the same way that classical solutions
to quantum field theories have done in the past.  We can address a
wide variety of problems, including
\begin{itemize}
\item
The stabilization of solitons by quantum effects in
theories that do not have classical soliton solutions.
\item
The direct calculation of induced charges, both fractional
and integer, carried by background field configurations.
\item
The analysis of the divergences and physical significance of
calculations of vacuum energies in the presence of boundaries --- {\it
i.e.\/} the traditional ``Casimir effect''.
\item
The computation of quantum fluctuations in  strong external fields.
\item
Quantum contributions to the properties of branes and domain walls.
\end{itemize}
This report presents an introduction to our methods and examples of
applications.  Sections 2--4 lay out the method and address
renormalization and computational efficiency.  Sections 5--9 introduce
examples.  In Section~2 we describe our method in general terms.  In
Section~3 we illustrate the method with the case of a charged boson
field in a bosonic background in three spatial dimensions.  We also
explain how to compute phase shifts and their Born approximations
efficiently.  In Section~4 we turn to renormalization using methods
adapted from dimensional regularization.  We work in $n$ space
dimensions and show that the leading terms in the Born expansion,
which diverge for integer $n$, can be unambiguously identified with
Feynman diagrams.  This approach resolves several longstanding
ambiguities in Casimir calculations.  In Section~5 we show how to
compute fractional and integer charges carried by background fields. 
We illustrate the importance of gauge invariant regularization of
Casimir calculations in a study of the charge carried by an
electrostatic ``hole'' in one space dimension.  In Section~6 we
consider a chiral model in one space dimension to show that quantum
effects of a heavy fermion can stabilize a soliton that is not present
in the classical theory.  We compute corrections to the energy and
central charge in $1+1$ dimensional supersymmetric models in
Section~7.  In Section~8 we extend our results to interfaces.  Finally
we conclude with a summary of some topics that are currently under
investigation or the subject for future projects.

\section{Overview}

We start with an overview of our method.  We treat the simplified
example of a fluctuating boson or fermion field of mass $m$ in a
static, spherically symmetric background potential $\chi(r)$ in three
dimensions.  Since we encounter divergences, we imagine that we have
analytically continued to values of the space dimension~$n$ where the
integrals are convergent.  In Section~4 we provide the rigorous
justification for this procedure.

We take the interaction Lagrangian ${\cal L}_{I}= g\overline\psi \chi
\psi$ for fermions and ${\cal L}_{I}= g\psi^\dagger \chi \psi$ for
bosons where $\psi$ is the fluctuating field.  We want to compute the
one--loop ``effective energy,'' the effective action per unit time. 
It is given either by the sum of all one loop diagrams with all
insertions of the background $\chi(r)$,
\be
\Delta E_{\rm bare}[\chi]
\hspace{-2.1cm}\parbox[b]{9.5cm}{ \raisebox{-1.5cm}{
\psfig{file=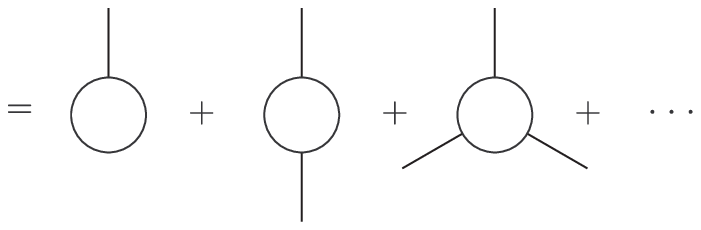,height=3cm,width=9cm}}}\quad ,
\label{insertions}
\ee
or by the ``Casimir sum'' of the shifts in the zero--point
energies of all the small oscillation modes in the background $\chi$,
\be
\Delta E_{\rm bare}[\chi] = \pm \frac{1}{2} \sum_j |\epsilon_j| -
|\epsilon^0_j|
\label{ebare0}
\ee
for bosons ($+$) and fermions ($-$) respectively.  Both of these
representations are divergent and require renormalization.  We start
from the second expression and work in the continuum.  We rewrite the
Casimir sum as a sum over bound states plus an integral over
scattering states, weighted by the density of states $\rho(k)$.  We
subtract from the integral the contribution of the trivial background,
which is given by the free density of states $\rho^0(k)$.  Thus we
have
\begin{equation}
\Delta E_{\rm bare}[\chi] = \pm\left(\frac{1}{2}\sum_j |\,\omega_j| +
\frac{1}{2} \int_0^\infty \omega(k) \left(\rho(k)-\rho^0(k)\right) dk \right)
\label{effengbare}
\end{equation}
 where $\omega_j$ denotes the energy of the $j^{\rm th}$ bound state,
 and $\omega(k)=\sqrt{k^2+m^2}$.

The density of states is related to the $S$-matrix and the phase
shifts by
\begin{equation}
	 \rho(k)- \rho^0(k) = \frac{1}{2\pi i}\frac{d}{dk}{\rm Tr}\ln
	 S(k) = \sum_\ell D^\ell \frac{1}{\pi}\frac{d\delta_\ell(k)}{dk}
	 \label{statedensity}
\end{equation}
where $\ell$ labels the basis of partial waves in which $S$ is
diagonal.  $D^{\ell}$ is the degeneracy factor.   For example,
$D^{\ell}=2\ell+1$ for a boson in three dimensions.  It is convenient
to use Levinson's theorem to express the  contribution of the
bound states to eq.~(\ref{effengbare}) in terms of their binding
energy. Levinson's theorem relates the number of bound states to the
difference of the phase shift at $k=0$ and $\infty$,
\begin{equation}
n_\ell^{\rm bound}= \frac{1}{\pi}( \delta_\ell(0) - \delta_\ell(\infty) )
=-\int_{0}^{\infty}dk\frac{d\delta_{\ell}(k)}{dk}\,.
\label{Levionson}
\end{equation}
Subtracting $m n_{\ell}^{\rm bound}$ from the sum over bound states in
eq.~(\ref{effengbare}) and using eqs.~(\ref{Levionson}) and
(\ref{statedensity}), we obtain

\begin{equation}
\Delta E_{\rm bare}[\chi] =
\pm\left(\frac{1}{2}\sum_{j,\ell}D^{\ell} (|\,\omega_{j,\ell}|-m) +
\int_0^\infty \frac{dk}{2\pi}\,  (\omega(k)-m)
 \sum_{\ell}D^{\ell}\frac{d\delta_\ell(k)}{dk} \right) 
 \label{ebare}
\end{equation}
where the sum over partial waves is to be performed before the $k$
integration.  While the phase shifts and bound state energies are
finite, $\Delta E_{\rm bare}[\chi]$ is divergent because the
$k$--integral diverges in the ultraviolet.  To better understand the
origin and character of the divergences, we go back to the
diagrammatic representation of the vacuum energy,
eq.~(\ref{insertions}).  Since we are working with a renormalizable
theory, only the first few diagrams are divergent, and these
divergences can be canceled by a finite number of counterterms.  The
series of diagrams gives an expansion of the effective energy in
powers of the background field $\chi(r)$.  Likewise, the phase shift
calculation can be expanded in powers of $\chi(r)$ using the Born
series,
\be
\delta^{N}_\ell(k) = \sum_{i=1}^N \delta^{(i)}_\ell(k)
\label{bseries}
\ee
where $\delta^{(i)}_{\ell}(k)$ is the contribution to the phase shift
at order $i$ in the potential $\chi(r)$.  In general, the Born
expansion is a poor approximation at small $k$, especially if the
potential has bound states, when it typically does not converge.  What
is important for us, however, is that the contributions to $\Delta
E_{\rm bare}$ from successive terms in the Born series correspond
exactly to the contributions from successive Feynman diagrams.  That
is, the $i^{\rm th}$ term in the Born series generates a contribution
to the vacuum energy which is exactly equal to the contribution of the
Feynman diagram with $i$ external insertions of $\chi$.

This correspondence is not trivial in light of divergences.  We have
verified the identification for the lowest order diagram by direct
comparison in $n$ space dimensions where both are finite.  At this
order the Born and Feynman contributions to $\Delta E_{\rm
bare}[\chi]$ are precisely equal as analytic functions of $n$ as we
will show in Section~4.  We have also performed various numerical
checks to verify the identification in higher orders.

\begin{figure}[t]
\centerline{\psfig{file=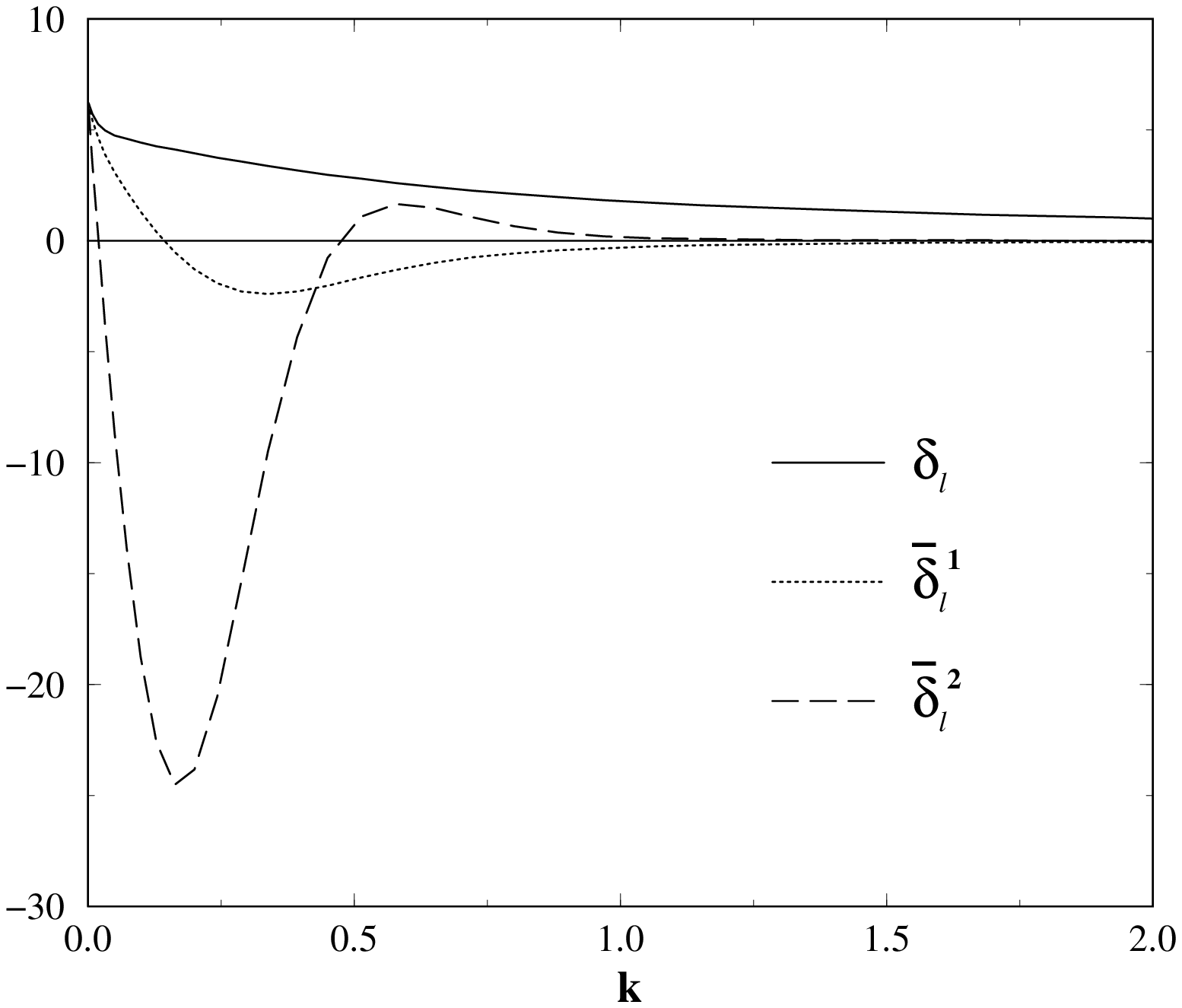,height=4.5cm,width=8.0cm}}
\fcaption{\label{fig_subt}
Typical phase shift in three dimensions, before and after subtracting
 the Born approximation.
}
\end{figure}

We then define the subtracted phase shift
\be
    \overline{\delta}^{\,N}_\ell(k)=\delta_\ell(k)-
     \delta^{\,N}_{\ell}(k)
\ee
where we take $N$ to be the number of divergent diagrams in the
expansion of eq.~(\ref{insertions}).
The effect of the Born subtraction is illustrated in
Fig.~\ref{fig_subt}.  Note that the subtracted phase shift
is large at small $k$, so the Born approximation is very
different from the true phase shift in this region.  However the Born
approximation becomes good at large $k$, so that the subtracted phase
shift vanishes quickly as $k\to\infty$.

Having subtracted the potentially divergent contributions to $\Delta
E_{\rm bare}[\chi]$ via the Born expansion, we add back in exactly the
same quantities as Feynman diagrams, $\sum_{i=1}^{N} \Gamma^{(i)}_{\rm
FD}[\chi]$.  We combine the contributions of the diagrams with those
from the counterterms, $\Delta E_{\rm CT}[\chi]$, and apply standard
perturbative renormalization conditions.  We have thus removed the
divergences from the computationally difficult part of the calculation
and re-expressed them as Feynman diagrams, where the regularization
and renormalization have been carried out with conventional methods. 
This approach to renormalization in strong external fields was first
introduced by Schwinger\cite{Schwinger} in his work on QED in strong
fields.  Combining the renormalized Feynman diagrams,
\begin{equation}
     \overline\Gamma^{N}_{\rm FD}[\chi]=
     \sum_{i=1}^{N} \Gamma^{(i)}_{\rm FD}[\chi]+\Delta E_{\rm CT}[\chi]
\end{equation}
with the subtracted phase shift calculation, we obtain the complete,
renormalized, one loop effective energy,
\bea
    \Delta E[\chi] &=&
    \pm \frac{1}{2} \sum_\ell D^\ell\left(\sum_j(|\,\omega_{j,\ell}|-m)+
    \int_0^\infty \frac{dk}{\pi} (\omega(k)-m)
    \frac{d}{dk}\overline{\delta}^{\,N}_\ell(k)\right)
    \nonumber \\ &&
    + \overline\Gamma^{N}_{\rm FD}[\chi]  
    \label{deltaE}
\eea
where the two pieces are now separately finite.  Since the $k$
integral is now convergent, we are free to interchange it with the sum
over partial waves or to integrate by parts.  This expression is
suitable for numerical computation, since it does not contain
differences of large numbers.  The massless limit is also smooth,
except for the case of one spatial dimension, where we expect
incurable infrared divergences\cite{Coleman1d}.

\section{An Example}

In this section we illustrate this approach with a simple example.  We
consider a charged scalar field, $\phi$, in a  classical scalar
background $\chi(r)$ in three dimensions\cite{3dscalar}.  We show how
to carry out a variational search for quantum--stabilized
nontopological solitons, which are local minima of the effective energy
functional, $E[\phi]$.  Although we do not find any solitons in this
simple model, we develop computational tools that will be useful when
we consider models with more complex structure.

\subsection{The model}

The model Lagrangian is given by
\bea
    {\cal L}&=&\frac{1}{2}(\partial_\mu \chi)^2
    -\frac{\lambda}{4!}\left(\chi^2-v^2\right)^2
    +\partial_\mu\phi^*\partial^\mu\phi-G\phi^*\chi^2\phi \cr
    &&\hspace{1cm} +\,a\,(\partial_\mu \chi)^2-b\left(\chi^2-v^2\right)
    -c\left(\chi^2-v^2\right)^2\, .
\label{toylag}
\eea
Here we have coupled $\phi$ to the square of the background field
$\chi$ so that the classical potential for $\chi$ is
positive--definite.  The potential for $\chi$ is of the
symmetry--breaking form with a minimum at $\chi=\pm v$.  By Derrick's
theorem, this model has no soliton at the classical level.

When we integrate out the dynamical $\phi$ field, we leave behind its
Casimir energy as a functional of $\chi$.  The divergences of the
Casimir energy are then canceled by the $\chi$-dependent counterterms
explicitly given in eq.~(\ref{toylag}).  The coefficients $a$, $b$ and
$c$ are determined by considering deviations $h=\chi-v$ from the
classical vacuum $\langle\chi\rangle=v$ and imposing the following
perturbative renormalization conditions:
\begin{itemize}
    \item[1)]
    the tadpole diagram with an external $h$ field vanishes,
    \item[2)]
    the location and residue of the pole of the $h$ propagator
    remain unchanged.
\end{itemize}
Condition~1) implies that there are no quantum corrections to the
vacuum expectation value $\langle\chi\rangle=v$.  Condition~2) gives
the standard on--shell renormalization conditions for $h$.  The small
oscillations potential for the $\phi$ fluctuations is
\be
V(r)=G\chi^2(r)-M^2=G(h^2(r)+2vh(r)) 
\label{toypot}
\ee
where $M=Gv^2$ is the mass of the $\phi$ field.

\subsection{Phase shifts and the Born approximation}

Next we compute the phase shifts in each partial wave.  An adaptation
of the variable phase method\cite{Ca67} provides a numerically stable
and efficient way to compute both the phase shifts and the first $N$
terms in the Born approximation for potentials $V(r)$ that fall faster
than $1/r^{2}$ as $r\to\infty$.

We write the Jost solution to the radial wave equation with angular
momentum $\ell$ as %
\be
    \varphi_\ell(k,r)=e^{2i\beta_\ell(k,r)} r h^{(1)}_\ell(kr)
    \label{defbeta}
\ee
where $h^{(1)}_\ell$ are Hankel functions describing outgoing
spherical waves.  The complex function $\beta_{\ell}(k,r)$ vanishes when
$V(r)=0$ and obeys the non--linear inhomogeneous ordinary differential
equation
\be
    -i\beta_\ell^{\prime\prime}-2ikp_\ell(kr)\beta^\prime_\ell
    +2(\beta_\ell^\prime)^2+\frac{1}{2}g V(r)=0 
     \label{eqbeta}
\ee
where a prime denotes differentiation with respect to $r$. The rational
function
\be
    p_\ell(x)=\frac{d}{dx} \ln\left(h^{(1)}_\ell(x)\right)
    \label{defpl}
\ee
can easily be computed numerically.  We have introduced a parameter
$g$ in eq.~(\ref{eqbeta}) to keep track of orders in the potential. 
Eventually we will set $g=1$.  With the boundary conditions
$\lim_{r\to\infty}\beta_{\ell}(k,r)=\beta_{\ell}^\prime(k,r)=0$, the
solution $u_{\ell}(k,r)$ that is regular at $r=0$ is a superposition
of $\varphi_{\ell}$ and $\varphi^{*}_{\ell}$ weighted by the
$S$-matrix $e^{2i\delta_{\ell}(k)}$,
\be
    u_{\ell} (k,r)=\varphi^*_\ell(k,r)
    +e^{2i\delta_\ell(k)}\varphi_\ell(k,r)\,.
     \label{scatsol}
\ee
Demanding that $u_{\ell}$ be regular at the origin determines the
phase shift,
\be
    \delta_\ell(k)=-\left.2{\rm Re}\,\beta_\ell(k,r)\right|_{r=0}\,.
    \label{phase1}
\ee

Next we extend this approach to the Born series, which is the
expansion of $\delta_{\ell}(k)$ in a power series in $g$.  From
eqs.~(\ref{bseries}) and (\ref{phase1}) we see that
\be
    \delta^{(i)}_\ell(k)=-2\left.{\rm Re}\,\beta^{(i)}_{\ell}(k,r)
    \right|_{r=0}
    \label{phaseborn}
\ee
where $\beta^{(i)}_{\ell}(k,r)$ are the terms in an expansion of
$\beta_{\ell}(k,r)$ in powers of $g$,
\begin{equation}
    \beta_{\ell}(k,r) \sim \sum_{i=1}^{\infty}g^{i}\beta^{(i)}_{\ell}(k,r)
    \, .
    \label{betaborn}
\end{equation}
We obtain a series of differential equations for the
$\{\beta^{(i)}_{\ell}\}$ by substituting the series expansion of
$\beta_{\ell}$ into eq.~(\ref{eqbeta}) and identifying powers of $g$,
\bea
-i\beta^{(1)\prime\prime}_{\ell}-2ikp_\ell(kr)\beta^{(1)\prime}_{\ell}
&=& - \frac{1}{2}V(r)
\nonumber \\
-i\beta^{(2)\prime\prime}_{\ell}-2ikp_\ell(kr)\beta^{(2)\prime}_{\ell}
&=&-2 (\beta^{(1)}_{\ell} )^2
\nonumber \\ &\vdots& \label{betaexp}
\eea
with the same boundary conditions as before.  The source term for each
successive order in the expansion involves only lower order terms. 
Thus these equations can be integrated simultaneously by integrating
the vector
$\{\beta_{\ell},\beta^{(1)}_{\ell},\beta^{(2)}_{\ell},\ldots\}$ in
from $r=\infty$.  The full phase shift and the Born approximations are
then determined by the value of the resulting vector at $r=0$.

At this point, we have collected all the ingredients for the continuum
part of the energy calculation.  From Levinson's theorem, we know how
many bound states to look for, and then the energies of the bound
states are easily obtained by standard shooting methods.

\subsection{Feynman diagrams}

\begin{figure}[t]
    \centerline{
    \psfig{file=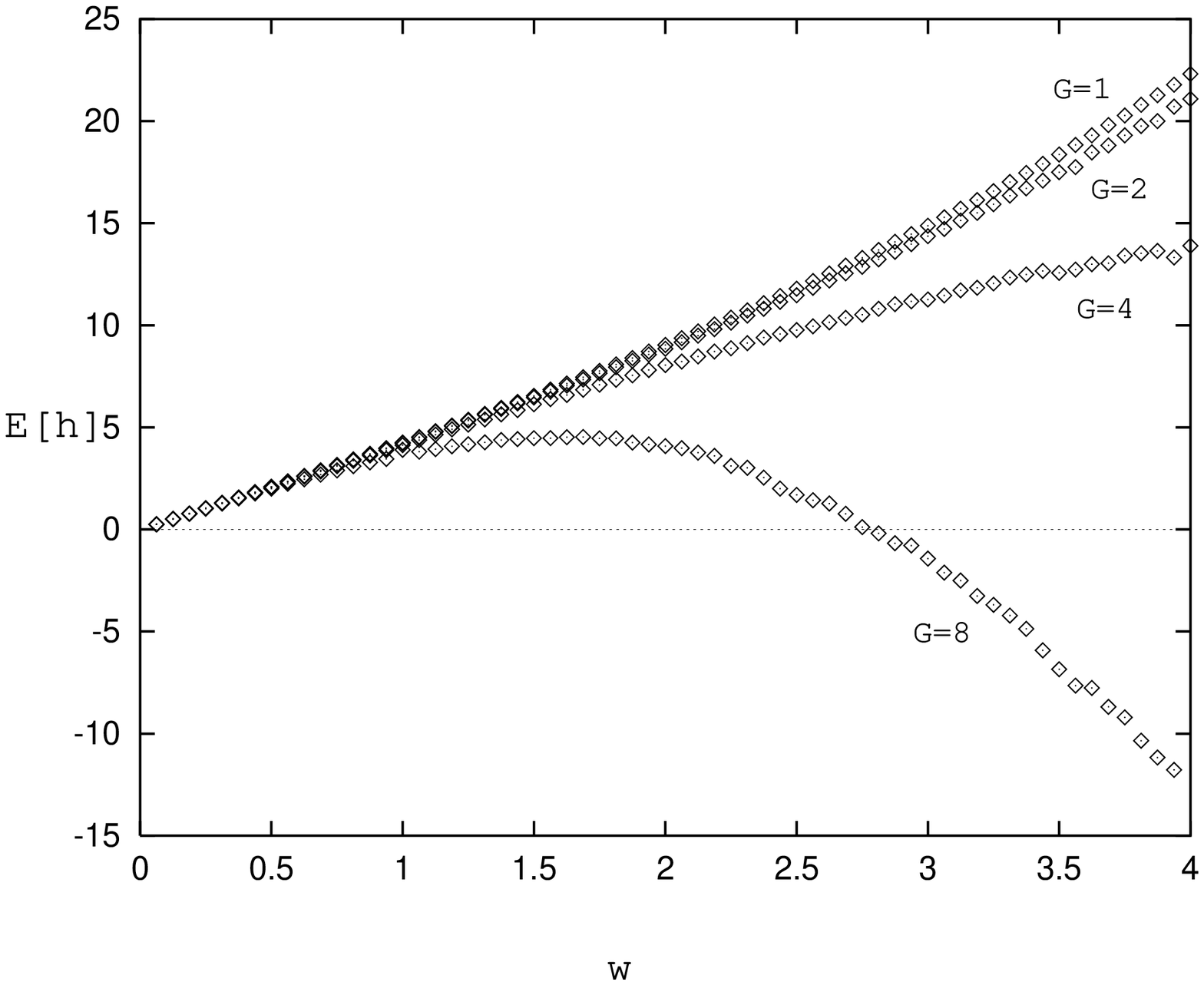,height=4.5cm,width=8.0cm}
    }
    \fcaption{\label{fig_toy}
    $E(d=1,w)$ in units of $v$ for $G=1,2,4$ and $8$ as function
    of $w$.}
\end{figure}

We now turn to the calculation of the Feynman diagrams, which we add
back into the energy functional to compensate for the Born
subtractions.  In this scalar model, only the diagrams with one and
two insertions of $V$ are divergent, so we take $N=2$ in
eq.~(\ref{deltaE}).  $\Gamma^{(1)}_{\rm FD}[h]$ is local and thus
completely canceled due to condition~1).  The divergences of
$\Gamma^{(2)}_{\rm FD}[h]$ are canceled by the $b$ and $c$ type
counterterms in eq.~(\ref{toylag}) leaving a renormalized
contribution, $\overline\Gamma^{(2)}_{\rm FD}[h]$.  The result is the
one--loop energy functional
\bea
    \Delta E[h]&=&  \overline\Gamma^{(2)}_{\rm FD}[\chi]
    +\sum_{j,\ell}(2\ell+1)(\omega_{j,\ell}-M) \cr
    &&-\int_0^\infty
    \frac{dk}{\pi}\frac{k}{\sqrt{k^2+M^2}}\sum_\ell(2\ell+1)            
    \left(\delta_\ell(k)-\delta^{(1)}_\ell(k)-
    \delta^{(2)}_\ell(k)\right)
    \label{etottoy}
\eea
where we have integrated by parts.  There is no surface term at
$k=\infty$ because the Born subtractions have removed the leading
large $k$ behavior of $\delta_{\ell}(k)$, and the Levinson subtraction
has eliminated the contribution from $k=0$.  There is also an overall
factor of 2 because the complex field $\phi$ contains two real
components.  The renormalized two--point Feynman diagram reads
\bea
    \overline\Gamma^{(2)}_{\rm FD}[\chi] &=&
    -\frac{4v^2G^2}{(4\pi)^2} \int_0^\infty
    \frac{q^2dq}{(2\pi)^2}q^2\tilde{h}^2 (q)\int_0^1 dx
    \frac{x(1-x)}{M^2-x(1-x)m^2}     \cr
    &&+ \frac{G^2}{(4\pi)^2} \int_0^\infty
    \frac{q^2dq}{(2\pi)^2}\tilde{V}^2(q)
    \int_0^1 dx \Big[ \ln \frac{M^2+x(1-x)q^2}{M^2-x(1-x)m^2} \cr
    &&\hspace{4cm}
    -\frac{x(1-x)m^2}{M^2-x(1-x)m^2}\Big]
    \label{Gamma2}
\eea
where $\tilde f(q)$ denotes the Fourier transform, $\tilde f(q) =\int
d^{3}r \exp(i\vec q\cdot\vec r) f(r)$, and $m=\sqrt{\lambda v^2/3}$ is
the mass of $h$.  We obtain the total energy functional by adding the
classical energy for $h$ to the one--loop quantum contribution,
\be
    E[h]=E_{\rm cl}[h]+\Delta E[h]  \,.
    \label{etotmod1}
\ee
To illustrate a typical variational search we introduce a two
parameter {\it ansatz}
\be
    h(r)=-dv e^{-r^2v^2/2w^2}  \,,
    \label{var1}
\ee
and compute the energy, $E(d,w)$, as a function of these variational
parameters.  Figure~\ref{fig_toy} shows the energy as a function of
$w$ at $d=1$ for several choices of $G$.
For fixed renormalized model parameters $G$ and $m$, we vary the {\it
ansatz\/} parameters $d$ and $w$.  One of the primary advantages of
our renormalization procedure is that it is manifestly independent of
the background field $\chi$, so we can confidently vary the {\it
ansatz\/} while keeping the model parameters fixed.  We can then see
if there exists a configuration for which the energy,
eq.~(\ref{etotmod1}), plus the energy required to explicitly occupy
the most tightly bound level is less than the mass of a free particle
in the trivial background.  If we find such a configuration, we know
that further variations can only lower the energy.  Thus there must
exist a stable soliton, which carries charge, but is too light to
decay into free particles.  In Ref.\cite{3dscalar} we did find such
configurations, but only for values of $G$ and $m$ where further
increase of $w$ yielded $E[h]<0$, signaling instability of the vacuum. 
This effect is depicted in the case of $G=8$ in Figure~\ref{fig_toy}. 
Although we did not find a non--trivial solution in this simple model,
we have demonstrated the practicality of the method in three
dimensions.

\section{Dimensional Regularization}

The identification of terms in the Born series with Feynman diagrams
is crucial to fixing the renormalization procedure precisely in our
approach.  No arbitrariness can be tolerated in the renormalization
process: if the manipulations of formally divergent quantities
introduce finite ambiguities, the method is useless.  There have been
controversies for many years concerning the proper renormalization
procedure for Casimir calculations.  Since we are studying
renormalizable quantum field theories, we know that the effective
energy can be calculated unambiguously.  In this section, we apply the
methods of dimensional regularization to scattering from a central
potential and prove that the lowest order term in the Born series is
equal to the lowest order Feynman diagram as an analytic function of
$n$, the number of space dimensions.  Since this is the most divergent
diagram --- quadratically divergent for $n=3$ --- we are confident
that the same method will regulate all other divergences in the
effective energy unambiguously.  For simplicity, we will consider the
self--interactions of a single real boson, $\phi(x)$.  The
generalization to fermions is discussed in Ref.\cite{tbaglevi}.

We consider a static, spherically symmetric background potential
$V(r)$ in $n$ dimensions.  For $n=1$, $V(r)$ reduces to a symmetric
potential with even and odd parity channels.  For $n\ne 1$ the
$S$-matrix is diagonal in the basis of the irreducible tensor
representations of $SO(n)$.  These are the traceless symmetric tensors
of rank $\ell$, where $\ell = 0,1,2,\ldots$.  We write the change in
the full density of states as a sum over partial waves
\begin{equation}
    \rho_n(k)-\rho_n^{0}(k)
    = \frac{1}{\pi} \frac{d}{dk}
    \sum_{\ell=0}^{\infty}D_{n}^{\ell} \delta_{n\ell}(k)
    \label{a1.3}
\end{equation}
where $D_{n}^{\ell}$ is the degeneracy of the $SO(n)$ representation
labeled by $\ell$.  For integer $n$ and $\ell$, $D_{n}^{\ell}$ is
given by the dimension of the space of symmetric, traceless tensors
with $\ell$ indices.  Replacing factorials by $\Gamma$-functions
allows us to define the analytic continuation of $D^{\ell}_{n}$,
\begin{equation}
    D_{n}^{\ell}=
    \frac{\Gamma(n+\ell-2)}{\Gamma(n-1)\Gamma(\ell+1)}(n+2\ell-2) \, .
    \label{a1.4}
\end{equation}
For $n=3$, $D_{n}^{\ell}$ reduces to $2\ell+1$ as expected.  As $n\to
1$ all the $D_{n}^{\ell}\to 0$ except for $\ell=0$ and $\ell=1$, for
which $\lim_{n\to 1}D^{0}_{n} = \lim_{n\to 1}D^{1}_{n} = 1$,
corresponding to the symmetric and antisymmetric channels
respectively.

The phase shifts are obtained by solving the radial Schr\"odinger
equation generalized to $n$ dimensions,
\begin{equation}
    -\psi''-\frac{n-1}{r}\psi' + \frac{\ell(\ell+n-2)}{r^{2}}\psi
    + V(r)\psi = k^{2}\psi
    \label{radschr}
\end{equation}
which reduces to Bessel's equation for $V=0$.  At the origin, the
regular solution $\psi_{n\ell}$ is proportional to $r^{\ell}$
independent of $n$.\footnote{Note that this produces the correct
wavefunction symmetry for $\ell=0$ and $\ell=1$ when $n=1$.}

Incoming and outgoing waves are given by generalized Hankel functions,
\begin{equation}
    h^{(1,2)}_{n\ell}(kr)=\frac{1}{(kr)^{\frac{n}{2}-1}}
    \left(J_{\frac{n}{2}+\ell-1}(kr)
    \pm i Y_{\frac{n}{2}+\ell-1}(kr)\right)
    \label{a1.6}
\end{equation}
and the phase shifts are defined in the usual way by writing the
solution $\psi_{n\ell}$ regular at the origin as
\begin{equation}
    \psi_{n\ell} \sim h^{(2)}_{n\ell}(kr)+e^{2i\delta_{n\ell}(k)}
    h^{(1)}_{n\ell}(kr)
    \label{a1.7}
\end{equation}
for large $r$, where the potential vanishes.  In $n$ dimensions,
the one--loop effective energy functional is
\begin{equation}
    \Delta E_n[\chi] = \frac{1}{2} \sum_{\ell=0}^{\infty}D_n^\ell
    (|\,\omega_{j,n\ell}| - m ) + \int_0^\infty\sum_j \frac{dk}{2\pi}
    (\omega(k)-m)\sum_{\ell=0}^{\infty}D_n^\ell \frac{d}{dk}
    \delta_{n\ell}(k)
    \label{cas1C}
\end{equation} 
where the $\omega_{j,n\ell}$ are the energies of the bound states in
each partial wave $\ell$.  Eq.~(\ref{cas1C}) is well defined for
$n<1$, where the integration over $k$ and the sum over $\ell$
converge.  We want to demonstrate explicitly that the contribution to
the energy from the first Born approximation to the phase shift is
precisely equal to the tadpole graph, which we calculate in ordinary
Feynman perturbation theory.  We show that these two quantities are
equal by computing both as analytic functions of $n$.

The first Born approximation to the phase shift is
\begin{equation}
    \delta_{n\ell}^{(1)}(k) = -\frac{\pi}{2}
    \int _0^\infty J_{\frac{n}{2} +\ell-1}(kr)^2 V(r) r\, dr
\label{1stBn}
\end{equation}
and its contribution to the Casimir energy is
\begin{equation}
    \Delta E^{(1)}_{n}[\chi\,]= \int_0^\infty \frac{dk}{2\pi}
    (\omega(k) - m)\sum_{\ell=0}^{\infty} D_\ell^n
    \frac{d\delta_{n\ell}^{(1)}(k)}{dk} \, .
\label{fborn}
\end{equation}
Using the Bessel function identity %
\begin{equation}
    \sum_{\ell=0}^{\infty}
    \frac{(2q+2\ell)\Gamma(2q+\ell)}{\Gamma(\ell+1)}J_{q+\ell}(z)^2 =
    \frac{\Gamma(2q+1)}{\Gamma(q+1)^2}
    \left(\frac{z}{2}\right)^{2q}
\label{BesselId}
\end{equation}
with $q=\frac{n}{2} - 1$, we sum over $\ell$ in eq.~(\ref{fborn}) and
obtain
\begin{equation}
    \Delta E^{(1)}_{n}[\chi\,]=
    \frac{\langle V \rangle (2-n) }{(4\pi)^\frac{n}{2}
    \Gamma\left(\frac{n}{2}\right)}
    \int_0^\infty (\omega(k)-m) k^{n-3}\, dk =
    \frac{\langle V \rangle m^{n-1}}{(4\pi)^\frac{n+1}{2}}
    \Gamma\Bigl(\frac{1-n}{2}\Bigr)
\label{DE1n}
\end{equation}
which converges for $0<n<1$. Here $\langle V\rangle$ is the $n$--dimensional
spatial average of $V(r)$,
\begin{equation}
    \langle V \rangle = \int V(x)\, d^nx =
    \frac{2\pi^\frac{n}{2}}{\Gamma\left(\frac{n}{2}\right)}
    \int_0^\infty V(r) r^{n-1}dr \, .
\end{equation}
The tadpole diagram is easily computed using dimensional
regularization, and the result agrees precisely with eq.~(\ref{DE1n}). 
Thus we can be certain that our method of subtracting the first Born
approximation and adding back the corresponding Feynman diagram is
correct.

\section{Fractional Induced Charges on Background Fields }

A scalar background configuration can carry the charge associated with
a boson or fermion field that fluctuates around it.  The charge can be
related to a trace of the Green's function and, in turn, to a sum over
bound states and an integral over phase shifts much like the
zero--point energy.  Our methods can be used to compute these charges. 
This approach was first developed and applied to simple cases by
Blankenbecler and Boyanovsky\cite{BB}.  Our methods provide
generalizations and applications to more complex systems.  Here we
derive the general result and present two non-trivial examples, the
electrostatic potential hole and the chiral bag in one dimension.

We consider a fermion field $\psi$ in a static background $V(r)$ in
$n$ dimensions.  Let $\omega$ denote the single--particle energies and
$\alpha$ the remaining quantum numbers.  The corresponding wave
functions are normalized so that
\be
	\int \psi^{\dagger}_\alpha(x,\omega) \psi^{}_{\alpha'}(x,\omega') \,
	d^nx = \delta(\omega - \omega') \delta_{\alpha , \alpha'}\,.
	\label{norm1}
\ee
For bound states the Dirac delta is replaced by a Kronecker
delta.  Then the charge density $j^{0}(x)$ is given by
\be
	j^0(x) = -\frac{1}{2} \sum_\alpha\, \int_{-\infty}^\infty
 	\hspace{-2.55em}\sum d\omega\,
	{\rm sign}(\omega) |\psi_\alpha(x,\omega)|^{2}  	
	\label{charge1}
\ee
where $\int \hspace{-0.85em} \Sigma d\omega\, (\ldots)$ denotes a sum
over bound states and integral over the continuum.  The Green's
function for the fermion field is
\be
	G(x,y,E) = \sum_\alpha\, \int_{-\infty}^\infty \hspace{-2.55em}
	\sum\quad \frac{d\omega}{2\pi}
	\frac{\psi_\alpha(x,\omega) \psi^{\dagger}_\alpha(y,\omega)}
	{E - \omega + i\, {\rm sign}(\omega) \epsilon}
	\label{green1}
\ee
and its imaginary part is the density of states,
	\be \rho(\omega)=\frac{dN}{d\omega} = {\rm Im~Tr} \:\frac{2}{\pi}
	\int G(x,x,\omega) \, d^nx = \frac{{\rm sign}(\omega)}{\pi}
	\sum_\alpha \int d^nx |\psi_\alpha(x,\omega)|^2\, .
	\label{green2}
\ee
Thus the charge $Q[V]=\int d^nx\, j^0(x)$ can be expressed via
the density of states. Combining eqs.~(\ref{charge1}) and
(\ref{green2}) and substituting eq.~(\ref{statedensity}) for the
density of states yields
\bea 	Q[V] &=& -\sum_\alpha \left( \int_m^\infty \frac{d\omega}{2\pi}
	\frac{d\delta_\alpha(\omega)}{d\omega} + \sum_{\omega_{\alpha j} >
	0} \frac{1}{2} -\int_{-\infty}^{-m} \frac{d\omega}{2\pi}
	\frac{d\delta_\alpha(\omega)}{d\omega} - \sum_{\omega_{\alpha j} <
	0} \frac{1}{2} \right)\cr &=& \frac{1}{2\pi} \sum_\alpha \left(
	\delta_\alpha(m) - \delta_\alpha(\infty) - \pi n^>_\alpha + \pi
	n^<_\alpha - \delta_\alpha(-m) + \delta_\alpha(-\infty) \right) 
	\label{main}
\eea
where we have defined the charge such that the configuration
$V\equiv0$ has zero charge.  Here $n^>_\alpha$ and $n^<_\alpha$ give
the number of bound states with positive and negative energy
respectively in each channel.

This result has an intuitive interpretation.  If a state leaves the
positive continuum but appears as a positive energy bound state, the
spectral asymmetry remains unchanged and the charge does not change. 
However, if this state crosses $\omega = 0$ and becomes a negative
energy bound state, the spectral asymmetry changes and the charge of
the configuration increases by one.  Likewise, if a negative energy
level moves up through zero energy, the charge decreases by one. 
Levinson's theorem tracks all states that enter and leave the two
continua at $\omega=\pm m$.  Thus
\be
 	\delta_\alpha(m) - \delta_\alpha(\infty) +
	\delta_\alpha(-m) - \delta_\alpha(-\infty) - \pi n_\alpha^< - \pi
	n_\alpha^> = 0\, ,
	\label{consistency}
\ee
so that\footnote{There are peculiarities in the symmetric channel in
one dimension.  See Ref.\cite{tbaglevi}}
\be
	Q = \frac{1}{\pi}\sum_\alpha \delta_\alpha(m) -
	\delta_\alpha(\infty) - \pi n_\alpha^>
	=\frac{1}{\pi}\sum_\alpha \pi n_\alpha^< - \delta_\alpha(-m) +
	\delta_\alpha(-\infty) \, .
	\label{main1}
\ee

\subsection{QED and the need for regularization}

Conserved charges are not renormalized.  That is, they do not receive
any contributions from the counterterms of the theory.  However, if
the theory is not regularized in a manner consistent with the symmetry
responsible for charge conservation, {\it ie.\/} gauge invariance,
then spurious, finite renormalization of the charge can occur.  The
case of QED in one dimension provides a clear illustration of this
problem.  Although the calculation of the charge is free of
divergences, proper attention to regularization is essential.

We consider an electrostatic potential hole with depth
$A^{0}\equiv\varphi$ and width $2L$.  The phase shifts in the two
parity channels are given by
\bea
    \frac{k}{m+\omega}\tan(kL+\delta^+(\omega))
    &=&\frac{q}{m+\omega+e\varphi}
    \tan qL \nonumber \\
    \frac{m+\omega}{k}\tan(kL+\delta^-(\omega))
    &=&\frac{m+\omega+e\varphi}{q}
    \tan qL\, ,
    \label{deltawell}
\eea
where $e$ is the elementary charge, $k=\sqrt{\omega^2-m^2}$ and $q
=\sqrt{(\omega+e\varphi)^2-m^2}$.  As $\omega\to\pm\infty$, the total
phase shift, $\delta^+ + \delta^-$, approaches $\pm2e\varphi L$ since
$q\to k\pm e\varphi$.  Substitution into eq.~(\ref{main1}) would
indicate a fractional induced charge on the potential hole.  Although
fractional charges are possible in other problems, as we will see
below, they do not occur in one dimensional electrostatics.  The
reason for this misleading result is that we have implicitly used a
regulator that is not gauge invariant.

To ensure that we maintain gauge invariance, we use dimensional
regularization and work in $n$ space dimensions, taking $n\to 1$ at
the end of the calculation.  The large $k$ behavior of the phase
shifts is given by the first Born approximation.  Using
eqs.~(\ref{1stBn}) and~(\ref{BesselId}), we find that in $n$
dimensions, the sum over all channels of the first Born approximation
to the phase shift is
\be
    \delta^{(1)}_n(\omega)=\omega k^{n-2}
    \frac{N_De\pi}{2^{n-2}\Gamma(\frac{n}{2})^2} \int_0^\infty A^0(r)
    r^{n-1} dr = \omega k^{n-2} \frac{N_D L^n e\varphi\pi}{2^{n-2} n
    \Gamma(\frac{n}{2})^2}
    \label{d1qed}
\ee
where $2N_D$ is the dimension of the Dirac algebra for spacetime
dimension $D=n+1$.  If we take the limit $n\to 1$ before taking the
limit $\omega\to\pm\infty$, we recover the previous fractional result
$\pm 2 e\varphi L$ with $N_D=1$.  However, the proper prescription is
to compute in $n<1$ dimensions and only take $n\to 1$ at the end.  In
that case we see that $\delta^{(1)}_n(\omega)$ vanishes as
$\omega\to\pm\infty$, so that the square well does not carry
fractional charge.  We note that other common schemes, such as
zeta-function regularization, would yield the spurious fractional
result.

Since this result involves only the first Born approximation, it can
be checked by considering the corresponding Feynman diagram.  The
lowest diagram contributing to the charge in an external field has two
insertions: one of the charge operator, $\gamma^{0}$, and the other of
the external field.  Thus we must consider the vacuum polarization
diagram, which in $D$ spacetime dimensions becomes
\bea
    i\Pi_{\mu\nu}(p)\hspace{-0.1cm}&=&\hspace{-0.1cm}-
    2e^2N_D \int_0^1\hspace{-0.15cm} d\xi
    \hspace{-0.1cm}
    \int \frac{d^Dk}{(2\pi)^D}
    \label{poltensor} \\ && \hspace{0.1cm}\times
    \frac{ 2\xi(1-\xi) (g_{\mu\nu}p^2 - p_\mu p_\nu) +
    g_{\mu\nu} \left[m^2-p^2 \xi(1-\xi) + k^2 (\frac{2}{D} - 1)\right] }
    {(k^2 + p^2 \xi(1-\xi) - m^2)^2}\, .
    \nonumber
\eea
If we had not regulated the theory by analytically continuing the
space-time dimension, we would not have found the last term, which
vanishes if we set $D=2$ from the outset.  Taking the $D\to 2$ limit
carefully shows that this term exactly cancels the two terms that
precede it, leaving the transverse form of the vacuum polarization
required by gauge invariance.  When these terms are not canceled, they
lead to the same fractional fermion number we obtained from the phase
shifts with the naive phase shift calculation.  Thus we must include
in our definition of the field theory the additional information that
the theory is regulated in order to preserve gauge invariance at the
quantum level.  Dimensional regularization provides a way to implement
this requirement in terms of both phase shifts and Feynman diagrams.

\subsection{Fractional charge in a chiral model}

As an example where genuine fractional charges occur, we consider a
chiral bag in $D=1+1$ dimensions.  The generalization to the three
dimensional bag model is discussed in Ref.\cite{tbaglevi}.

We consider a free fermion on the half--line $x>0$ satisfying
the boundary condition
\be
    ie^{i\gamma_5 \theta} \Psi = \gamma^1 \Psi
    \label{boundary1d}
\ee
at $x=0$, with $-\frac{\pi}{2}\leq\theta\leq\frac{\pi}{2}$
parameterizing the model.  We break the scattering into generalized
parity channels.  The corresponding phase shifts are given by
\be
    \cot \delta^+(\omega) = -\frac{k}{\omega - m} \tan \beta
    \quad{\rm and}\quad
    \tan \delta^-(\omega) = \frac{k}{\omega - m} \tan \beta
\label{phase1dbag}
\ee
where $\beta=\frac{\pi}{4}-\frac{\theta}{2}$.  At the thresholds
$\omega\to\pm m$, the phase shifts are either 0 or $\frac{\pi}{2}$. At
large $\omega$, we find $\delta^+(\pm\omega) + \delta^-(\pm\omega)\to
-\frac{\pi}{2}\pm(\frac{\pi}{2}-\theta)$. There is a single bound
state determined by the condition
\begin{equation}
    \sqrt{m^2-\omega^2}=(m+\omega)\tan\beta\,.
\end{equation}
Collecting these results into eq.~(\ref{main1}), we find that the
fermion number is $\frac{\theta}{\pi} - \frac{{\rm sign}
(\theta)}{\pi}$.  We have fixed the overall integer constant by noting
that the jump in fermion number as a function of $\theta$ should occur
where there is a Jackiw-Rebbi\cite{JR} zero mode, namely at
$\theta=0$.  One can also obtain this constant by considering the
fermion number as a function of temperature\cite{DunneT}.

\section{Chiral Model in One Dimension}

As an application of our method, we show how a quantum soliton can
appear in a theory with a heavy fermion.  We consider a
one--dimensional chiral model in which the fermion gets its mass from
its coupling to a scalar condensate.  It is easy to find a spatially
varying scalar background which has a tightly bound fermion level.  If
the classical energy of the background field plus the energy of the
tightly bound fermion is less than the free fermion mass $m$, this
configuration would appear to be a stable soliton, since it is unable
to decay into free fermions.  However, the energy of the lowest
fermion level enters at the same order in $\hbar$ as the full
one--loop fermion effective energy, since the latter simply
corresponds to the shift of the zero--point energies,
eq.~(\ref{ebare0}), of all the fermion modes.  The question of
stability can therefore only be addressed by computing the full one
loop effective energy.  Here we summarize our analysis of this system
and show that it supports stable solitons.  We also illustrate how our
scattering theory methods generalize to fermions.  More details of
this calculation can be found in Ref.~\cite{heavyfermion}.

\subsection{The model}

We consider a chiral model in one dimension with a symmetry--breaking
scalar potential.  We couple a two--component real boson field
$\vec{\phi}=(\phi_1,\phi_2)$ chirally to a fermion $\Psi$ %
\begin{eqnarray}
    {\cal L}=\frac{1}{2}\,
    \partial_\mu\vec{\phi}\cdot\partial^\mu\vec{\phi}
    -V(\vec{\phi}\,)
    +\bar{\Psi}\left\{i\partial \hskip -0.55em /
    - G\left(\phi_1+i\gamma_5\phi_2\right)     \right\}\Psi
    \label{lagd11}
\end{eqnarray}
where the potential for the boson field is given by
\begin{eqnarray}
    V(\vec{\phi})=\frac{\lambda}{8}
    \left[\vec{\phi}\cdot\vec{\phi}
    -v^2+\frac{2\alpha v^2}{\lambda}\right]^2
    -\alpha v^3\left(\phi_1-v\right)+{\rm const.}
    \label{potential}
\end{eqnarray}
$V(\vec\phi\,)$ has its minimum at $\vec\phi=(v,0)$.  Terms
proportional to $\alpha$ break the chiral symmetry explicitly.  If
$\alpha$ we set to zero, the chiral symmetry appears to
break spontaneously, but quantum fluctuations in one dimension restore
the symmetry\cite{Coleman1d}.  For large enough $\alpha$, the classical
vacuum $\vec\phi=(v,0)$ is stable against quantum corrections and
$m=Gv$ is the fermion mass. The coefficient $c$ in the
counterterm Lagrangian \begin{equation}
{\cal L}_{\rm c.t.}=c\left(\vec{\phi\,}\cdot\vec{\phi\,}-v^2\right)
\label{Lct1p1}
\end{equation}
is fixed by the condition that the quantum corrections do not change
the VEV of $\vec{\phi\,}$.  This model has no stable soliton solutions
at the classical level.

We are interested in the mass of the lightest state carrying unit
fermion number.  If its mass is less than $m$, this state is a stable
soliton.  We neglect boson loops, so that the effective energy is
given by the sum of the classical and the fermion loop contributions,
$E_{\rm tot}[\vec\phi\,]=E_{\rm cl}[\vec\phi\,]+E_{\rm
f}[\vec\phi\,]$.  This approximation is exact in the limit where the
number of independent fermion species becomes large.  The fermion
contribution to the effective energy is $E_{\rm f}=E_{\rm Cas}+E_{\rm
val}$ where $E_{\rm Cas}$ is the sum over zero--point energies,
calculated with the methods we have developed.  $E_{\rm val}$ is the
energy required for the soliton to have unit charge.  Using the
methods of the previous section, we can calculate the fermion number
of the background field.  If a level has crossed zero, then the
background field will already carry the required fermion number and
$E_{\rm val}=0$.  If the background field has zero charge, we must
explicitly fill the most tightly bound level, giving $E_{\rm val} =
\epsilon_0$, where $\epsilon_0$ is the energy of that level.

\subsection{Phase shifts}

We consider configurations with $\phi_1(x)=\phi_1(-x)$ and
$\phi_2(x)=-\phi_2(-x)$, so that parity is a good quantum number. 
Hence the solutions $\psi(x)$ to the Dirac equation in the background
$\vec \phi\,$ can be decomposed into parity channels,
$\gamma^0\psi_\pm(-x)= \pm\psi_\pm(x)$.  Choosing the Majorana basis
($\gamma^0=\sigma_2$, $\gamma^1=i\sigma_3$, $\gamma_5=\sigma_1$), we
have
\begin{equation}
    \psi_\pm(0)  \propto \pmatrix{1\cr\pm i}\,.
    \label{parity}
\end{equation}
Then the solution to the Dirac equation is
\begin{eqnarray}
    \varphi_{+k}(x)&=&\pmatrix{f(x) \cr
     i\frac{f^\prime(x)+G\phi_1(x)f(x)}{\omega+G\phi_2(x)}} \cr \cr
    \varphi_{-k}(x)&=&\pmatrix{f^*(x) \cr
     i\frac{f^{*\prime}(x)+G\phi_1(x)f^*(x)}{\omega+G\phi_2(x)}} 
    \label{majspin}
\end{eqnarray}
where $f(x)$ obeys a real second--order differential equation, which
is amenable to the variable phase method.  We write $f(x)=
e^{i\beta(x,\omega)} e^{ikx}$ with boundary conditions
$\beta(\infty,\omega)=\beta^\prime(\infty,\omega)=0$, so that
$\varphi_{+k}$ and $\varphi_{-k}$ asymptotically describe outgoing and
incoming plane wave spinors, respectively.  We substitute into the
Dirac equation and solve the resulting differential equation for
$\beta(x,\omega)$ numerically,
\begin{eqnarray}
&&-i\beta''(x,\omega)+2k\beta'(x,\omega)+\beta^{\prime2}(x,\omega) - m^2 +
G^2\phi_1^2(x) + G^2\phi_2^2(x) - G\phi_1'(x)
\cr
&&+\frac{G\phi_2'(x)}{\omega + G\phi_2(x)}
\bigl[G\phi_1(x)+i(k+\beta'(x,\omega))\bigr] = 0 \,.
\hspace{1cm}
\label{eqbeta1}
\end{eqnarray}
We define the phase shifts $\delta_\pm(\omega)$ by writing 
scattering wavefunctions in the basis of parity eigenstates,
\begin{equation}
    \psi_{\pm}(x)=\varphi_{-k}(x)\pm\frac{m-ik}{\omega}\,
    e^{2i\delta_\pm(\omega)}\,\varphi_{+k}(x)
    \label{Diracphase}
\end{equation}
where we have introduced the factor $(m-ik)/\omega$ to guarantee that
$\delta_\pm=0$ when $\vec{\phi}=(v,0)$.  Imposing the boundary
conditions eq.~(\ref{parity}) onto the scattering solution
eq.~(\ref{Diracphase}) yields
\begin{equation}
    \delta_{\pm}(\omega) = -\mathop{\rm Re} \beta (0,\omega)-
    \arg\Bigl[1+\frac{i\beta^\prime(0,\omega) + G(\phi_1(0) - v)}
    {\mp\omega+m+ik}\Bigr]\,.
    \label{beta}
\end{equation}
Finally we sum the contributions from positive and negative energies
and both parities,
\begin{equation}
    \delta_F(k)=\delta_+(\omega(k))+\delta_+(-\omega(k))
    +\delta_-(\omega(k))+\delta_-(-\omega(k))\, .
    \label{totdel}
\end{equation}

Renormalization is particularly simple in this model.  The first and
second Born approximations corresponding to the Feynman diagrams with
one and two insertions of $[\vec{\phi}-(v,0)]\,$ diverge.  However,
the divergences are related by chiral symmetry.  Both are canceled by
a counterterm proportional to $\vec\phi\,^{2}-v^{2}$.  It suffices to
subtract the first Born approximation to $\delta(k)$ and the part of
the second related to it by chiral symmetry,
\begin{equation}
    \delta^{(1)}(k)=\frac{2G^2}{k}\int_0^\infty dx
     \left(v^2-\vec{\phi}\,^2(x)\right)\, .
    \label{subtr}
\end{equation}
The condition that the VEV of $\vec\phi\,$ does not get renormalized
requires that the counterterm exactly cancel the Feynman diagrams that
are added back in compensation for the Born subtractions.  Thus we
have
\begin{eqnarray}
    E_{\rm Cas}=-\frac{1}{2}\sum_j\left(
    \left|\,\omega_j\right|-m\right)
    -\int_0^\infty\frac{dk}{2\pi}\, \left(\omega(k)-m\right)
    \frac{d}{dk}\left(\delta_{\rm F}(k)-\delta^{(1)}(k)\right)\, .
\label{efermion}
\end{eqnarray}

\subsection{Numerical studies}

\begin{figure}[hbt]
\centerline{
\psfig{file=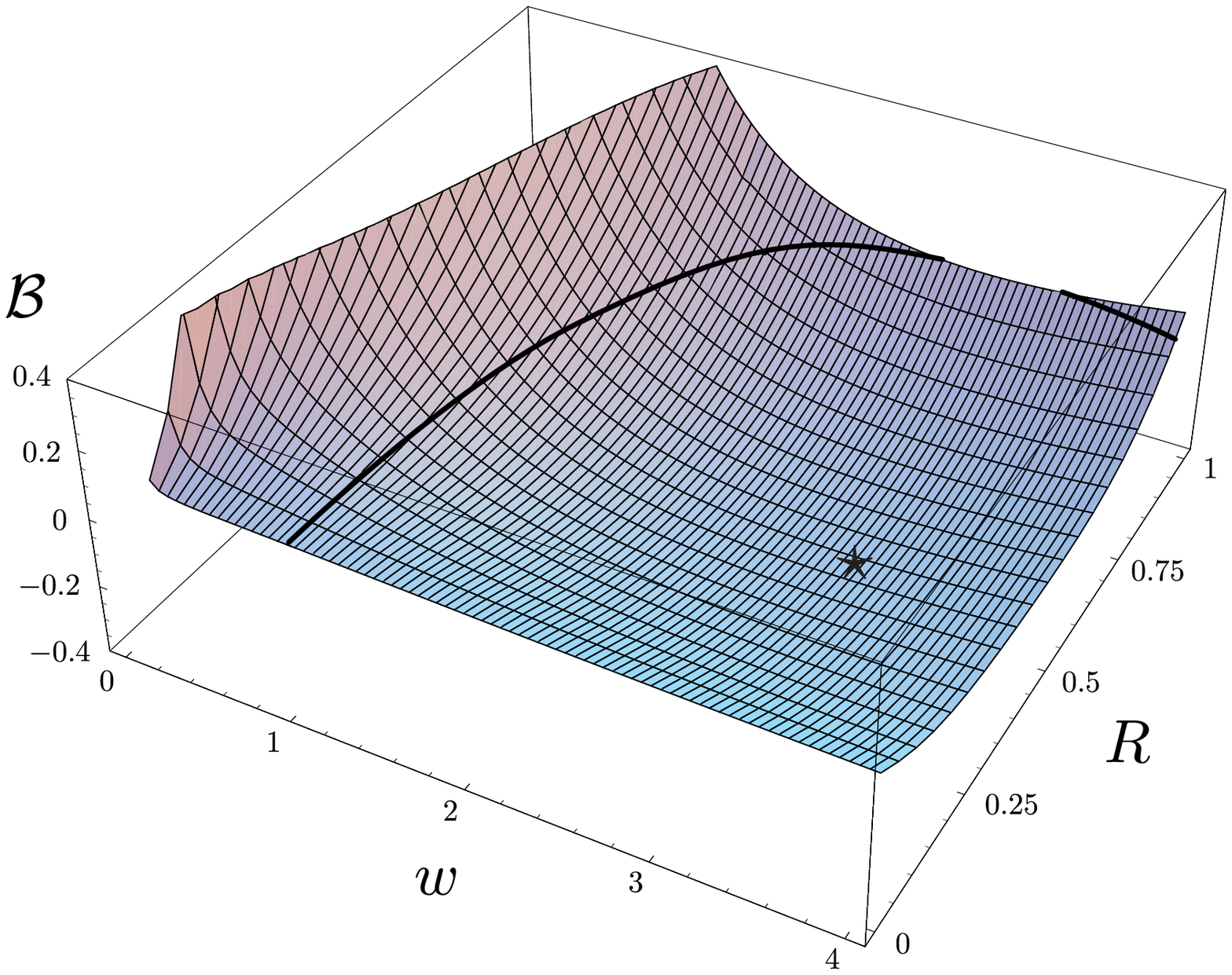,height=4cm,width=6cm}}
\fcaption{${\cal B}$ as a function of the \emph{ansatz} parameters
for the class of \emph{model} parameters characterized by the
relations $\alpha = 0.5G^2$, $\tilde \lambda = G^2$, and $v = 0.375$. 
A solid curve marks the contour ${\cal B} = 0$.  The star indicates
the minimum at $w=2.808$ and $R=0.586$.}
\label{example}
\end{figure}

\begin{figure}[hbt]
\centerline{
\psfig{file=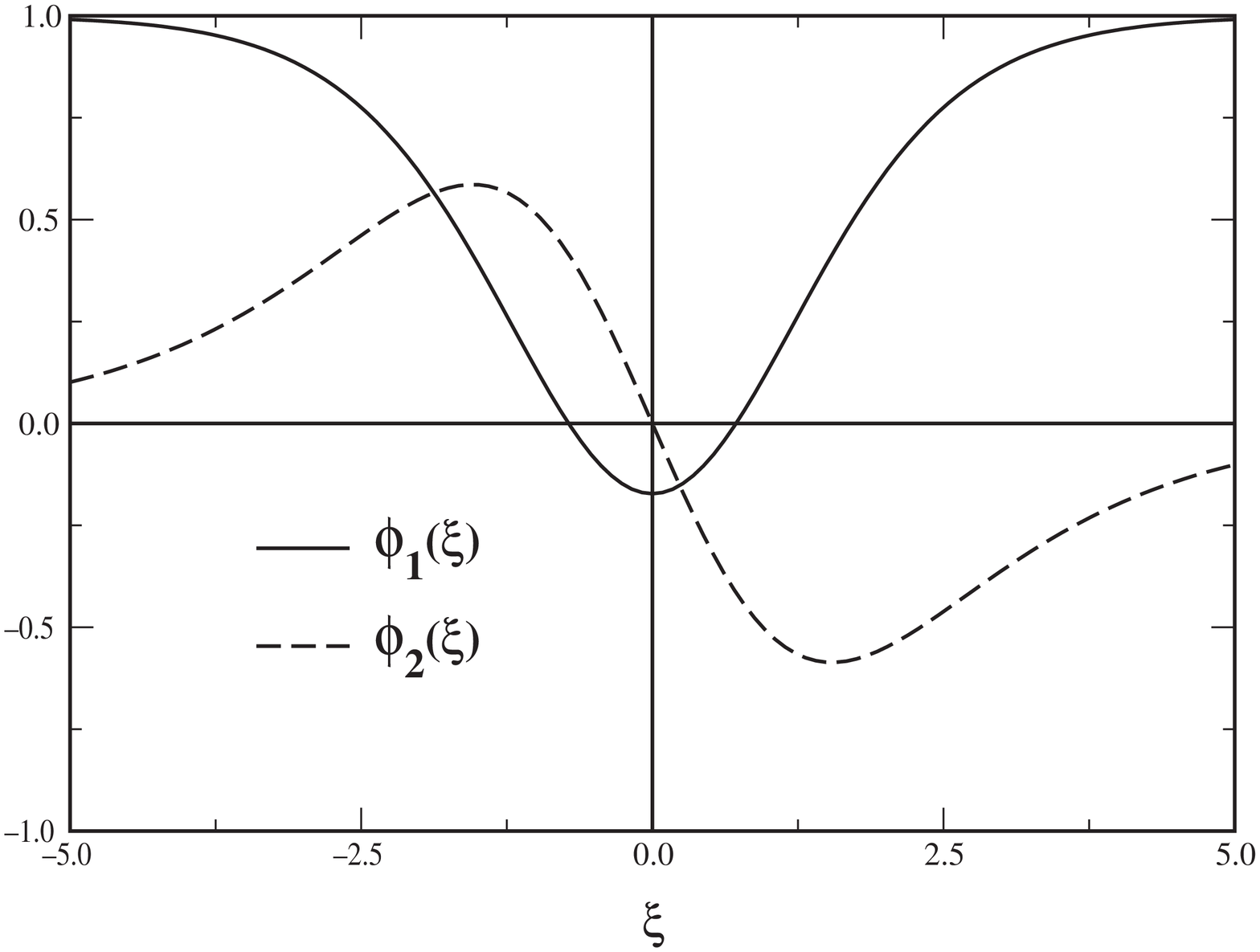,height=4cm,width=6cm} \quad
\psfig{file=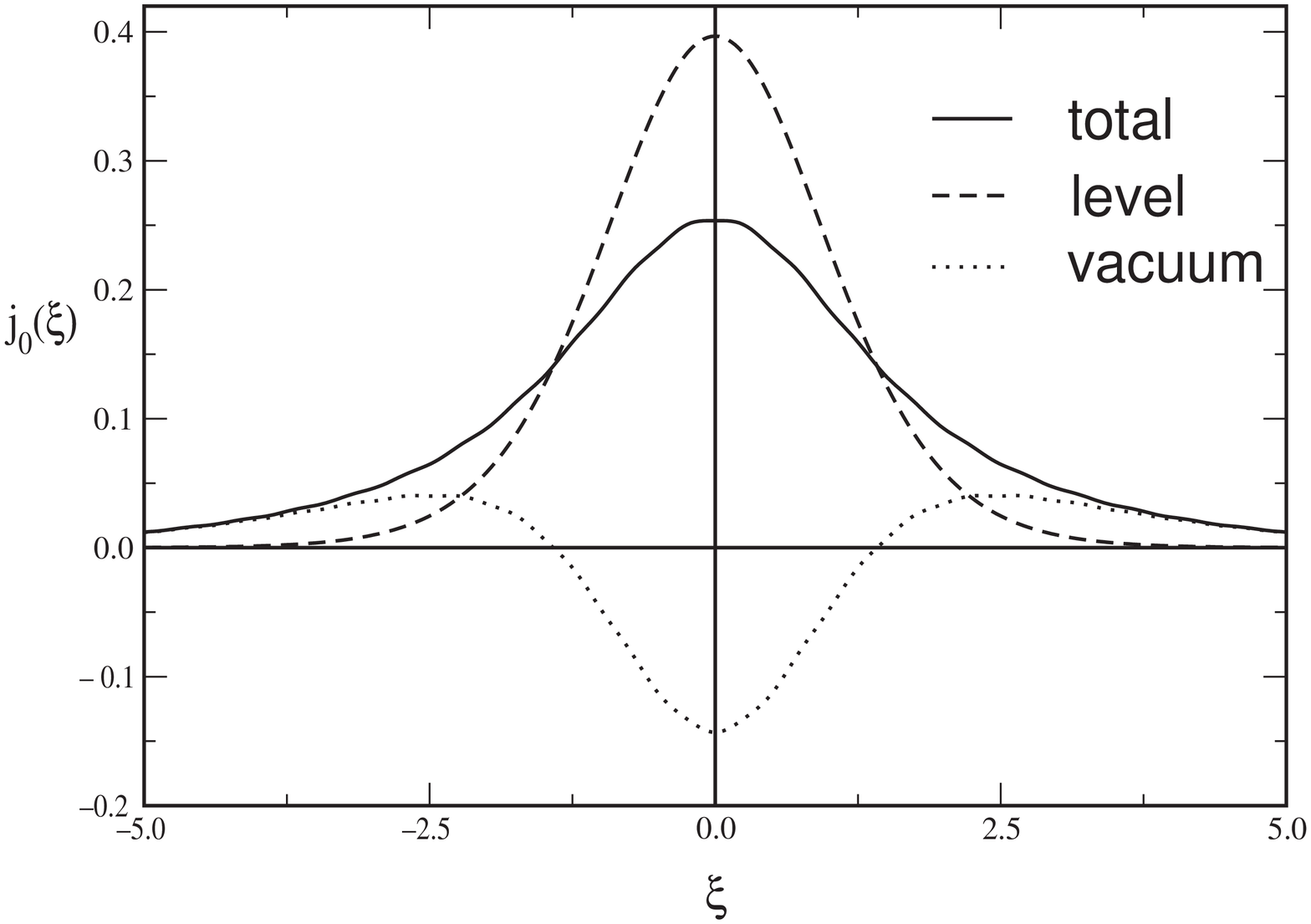,height=4cm,width=6cm}}
\vskip0.5cm
\fcaption{\label{config_fig} $\phi_{1}$, $\phi_2$, and the fermion
number density $j_0$ at the variational minimum.  The left panel shows
$\phi_1(\xi)$ and $\phi_2(\xi)$, and the right panel shows the charge
density $j_0(\xi)$, which gets contributions from both the polarized
fermion vacuum eq.~(\protect\ref{charge1}) and the filled valence
level.  The model parameters are as in Figure~\protect\ref{example}.}
\label{soliton}
\end{figure}

We consider variational {\it ans\"atze} for the background field.  As
$x\to\pm\infty$, $\vec\phi\, $ must go to its vacuum value, $(v,0)$.
We  find that energetically favored configurations execute a loop in the
$(\phi_{1},\phi_{2})$ with radius $R>v$ so that they enclose the
origin.  A simple {\it ansatz\/} with these properties is
\begin{equation}
    \phi_1+i\phi_2=v\left\{1-R+R\,\exp
    \left[i\pi\left(1+\tanh(Gvx/w)\right)\right]\right\}
    \label{variation}
\end{equation}
with the width ($w$) and amplitude ($R$) as variational parameters. 
For particular model parameters $G$, $\alpha$, $\lambda$ and $v$, we
compute ${\cal B}=E_{\rm tot}/m-1$ as a function of the variational
parameters $w$ and $R$.  We show the resulting binding energy surface
in Figure~\ref{example} for one set of model parameters.  The contour
${\cal B}=0$ separates the region in which the effective energy of
background configuration is less than $m$ from the region in which it
is larger than $m$.  The maximal binding is indicated by a star.  In
Figure~\ref{config_fig} we present the profiles $\phi_1$ and $\phi_2$
corresponding to this variational minimum as functions of the
dimensionless coordinate $\xi=xm$.
 This background field configuration does not carry
fermion number in this case, so the most strongly bound level must
explicitly be occupied. The total charge density is shown in
Figure~\ref{config_fig}.  It receives contributions from the polarized
fermion vacuum, given by eq.(\ref{charge1}), and from the explicitly
occupied valence level, given by $\psi_0^\dagger(x)\psi_0(x)$ where
$\psi_0(x)$ is the bound state wavefunction of the valence level.

\begin{figure}[tb]
    \centerline{\psfig{figure=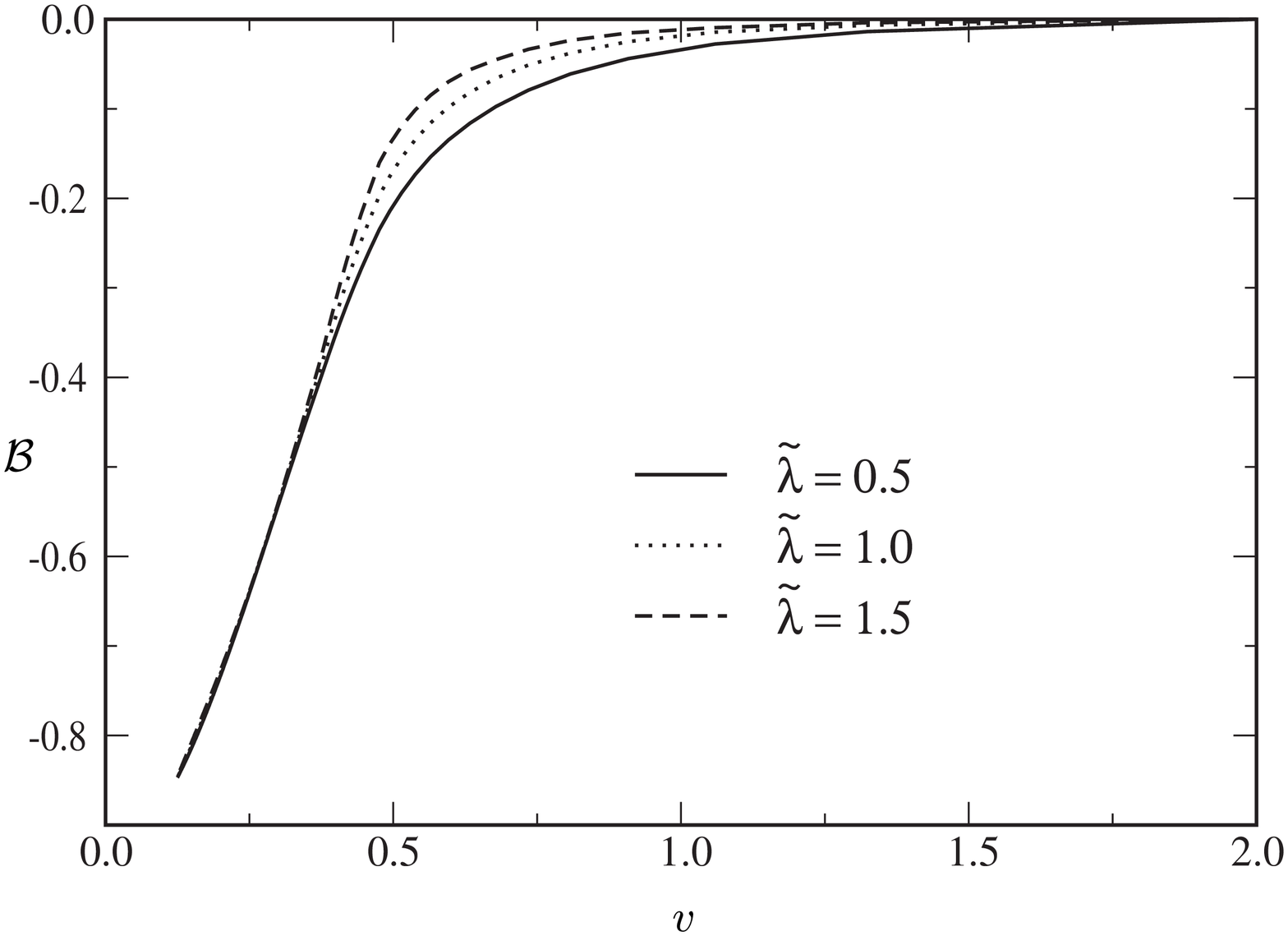,height=5.0cm,width=5.5cm}
    \hspace{1cm} \psfig{figure=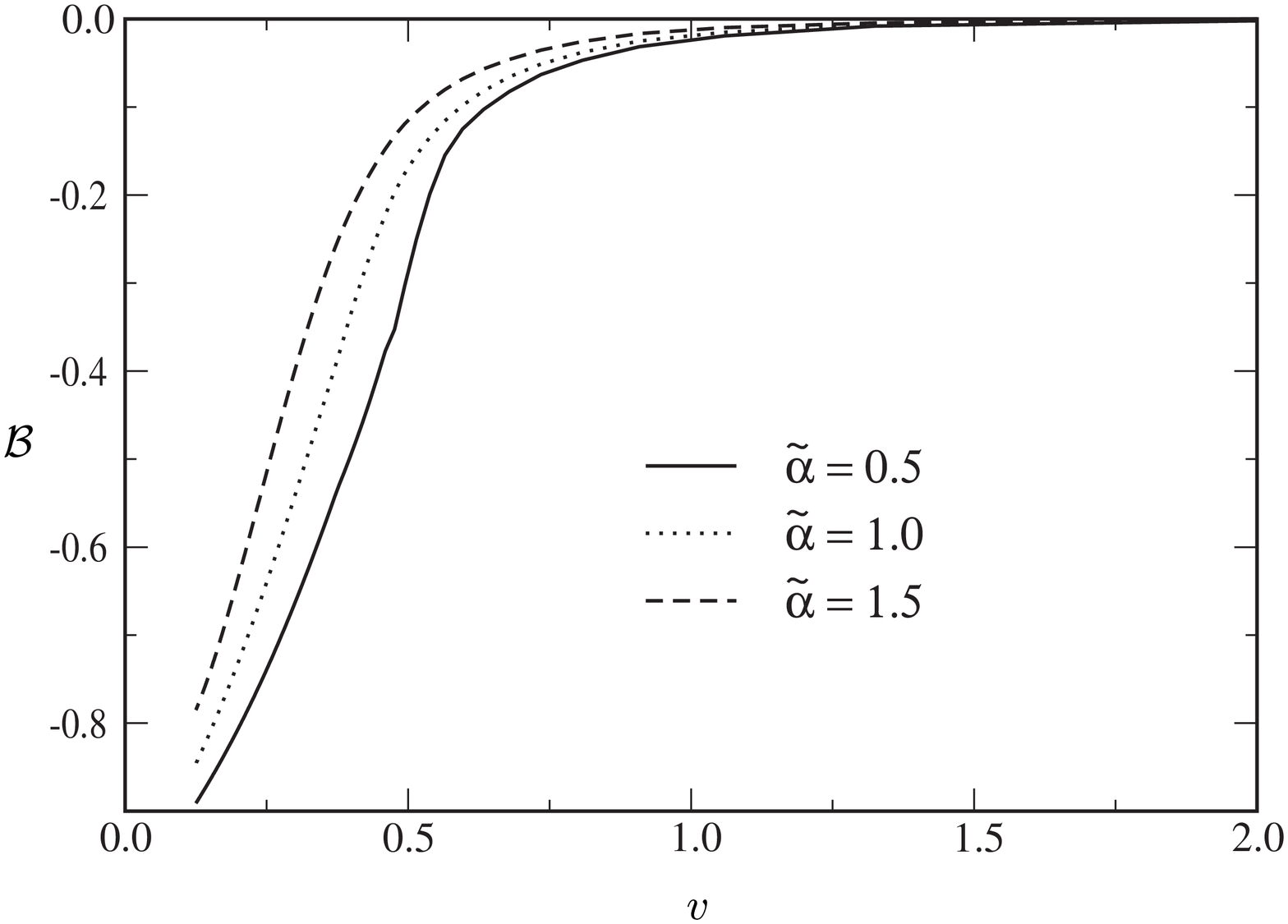,height=5.0cm,width=5.5cm} }
    \fcaption{\label{fig_1}
    The maximal binding energy as a function of
    the model parameters as obtained from the {\it ansatz}
    eq.~(\protect\ref{variation}) in units of $m$.  The dimensionless
    parameters are defined by $\tilde{\alpha}={\alpha}/{G^2}$ and
    $\tilde{\lambda}={\lambda}/{G^2}$.}
\end{figure}

Figure~\ref{fig_1} shows the result of repeating the binding energy
calculation for  various sets of model parameters.
When ${\cal B}$ is negative, the configuration is a fermion with lower
energy than a fermion propagating in the trivial background.  Since
the true minimum of the energy will have even lower energy, we know
that a soliton exists.

We have extended this analysis to a chiral Yukawa model with $SU(2)$
symmetry in three dimensions.  The analysis is more complicated:
rotational symmetry is replaced by grand spin, the sum of rotations in
spatial $SU(2)$ and isospin $SU(2)$, and diagrams up to fourth
order in the external field are divergent.  Nevertheless, the program
can still be carried out\cite{3dpaper}.  However, we do not find
evidence for a bound fermionic soliton in this theory.  In general,
binding is weaker than in one dimension and it occurs in regions of
parameter space where the model or the restriction to one fermion loop
is internally inconsistent. It is possible that expanding the model
to also include gauge fields may change this result, and work is
underway to consider this possibility.

\newpage
\section{Quantum Corrections to the Energy and Central Charge for
Supersymmetric Solitons in 1+1  Dimensions}

$N=1$ supersymmetric models in $1+1$ dimension provide a simple
example of our methods.  The ability to study configurations that are
not solutions to the equations of motion (and therefore not
supersymmetric) and to handle renormalization unambiguously allows us
to resolve long-standing questions regarding saturation of the BPS
bound\cite{many,vanN}.  Here we present only a brief introduction to
the results of Ref.\cite{1d} and refer the reader there for a more
complete presentation.

We consider the Lagrangian
\begin{equation}
  {\cal L} = \frac{m^2}{2\lambda} \left((\partial_\mu \phi)
  (\partial^\mu \phi) - U(\phi)^2 +i\bar\Psi
  \partial\hspace{-0.53em}\slash \Psi - U^\prime(\phi) \bar \Psi \Psi
  \right)
  \label{SUSYlag}
\end{equation}
where $\phi$ is a real scalar, $\Psi$ is a Majorana fermion, and
$U(\phi) = W'(\phi)$ where $W(\phi)$ is the superpotential. If
$U(\phi)^{2}$ is of the symmetry breaking form with equal minima at
$\phi=\pm 1$,  then a soliton is a solution to
\be
    \frac{d\phi_0(x)}{dx}= -U(\phi_0(x))
    \label{kink}
\ee
where $\phi_{0}\to\pm 1$ as $x\to\pm\infty$.  An antisoliton is
obtained by sending $x$ to $-x$.  The boson and fermion small
oscillation modes are given by
\begin{eqnarray}
  \left( -\frac{d^2}{dx^2} + U^\prime(\phi_0)^2 +
  U(\phi_0)U^{\prime\prime}(\phi_0) \right) \eta_k(x) &=& \omega^2
  \eta_k(x) \\ \gamma^0 \left(-i\gamma^1 \frac{d}{dx} +
  U^\prime(\phi_0)\right) \psi_k(x) &=& \omega \psi_k(x).   
  \label{Diraceq}
\end{eqnarray}
Defining
\begin{eqnarray}
    V(x) &=& U^\prime(\phi_0)^2 + U(\phi_0)U^{\prime\prime}(\phi_0) - m^2
    \cr \tilde V(x) &=& U^\prime(\phi_0)^2 -
    U(\phi_0)U^{\prime\prime}(\phi_0) - m^2
\end{eqnarray}
and squaring the Dirac equation, we obtain
\begin{eqnarray}
      \left( -\frac{d^2}{dx^2} + V(x) \right) \eta_k(x) &=&   \omega^2
       \eta_k(x) \cr        \pmatrix{-\frac{d^2}{dx^2}+V(x) & 0 \cr
       0 &-\frac{d^2}{dx^2}+\tilde{V}(x)}\psi_k(x) &=& k^2\psi_k(x) \,.
       \label{susydir}
\end{eqnarray}
It is easy to show that the bound state spectra of the
effective scalar potentials $V(x)$ and $\tilde V(x)$ will always
coincide, except possibly for zero modes.

\subsection{Supersymmetric Spectrum}

\begin{figure}[hbtf]
\centerline{
\psfig{file=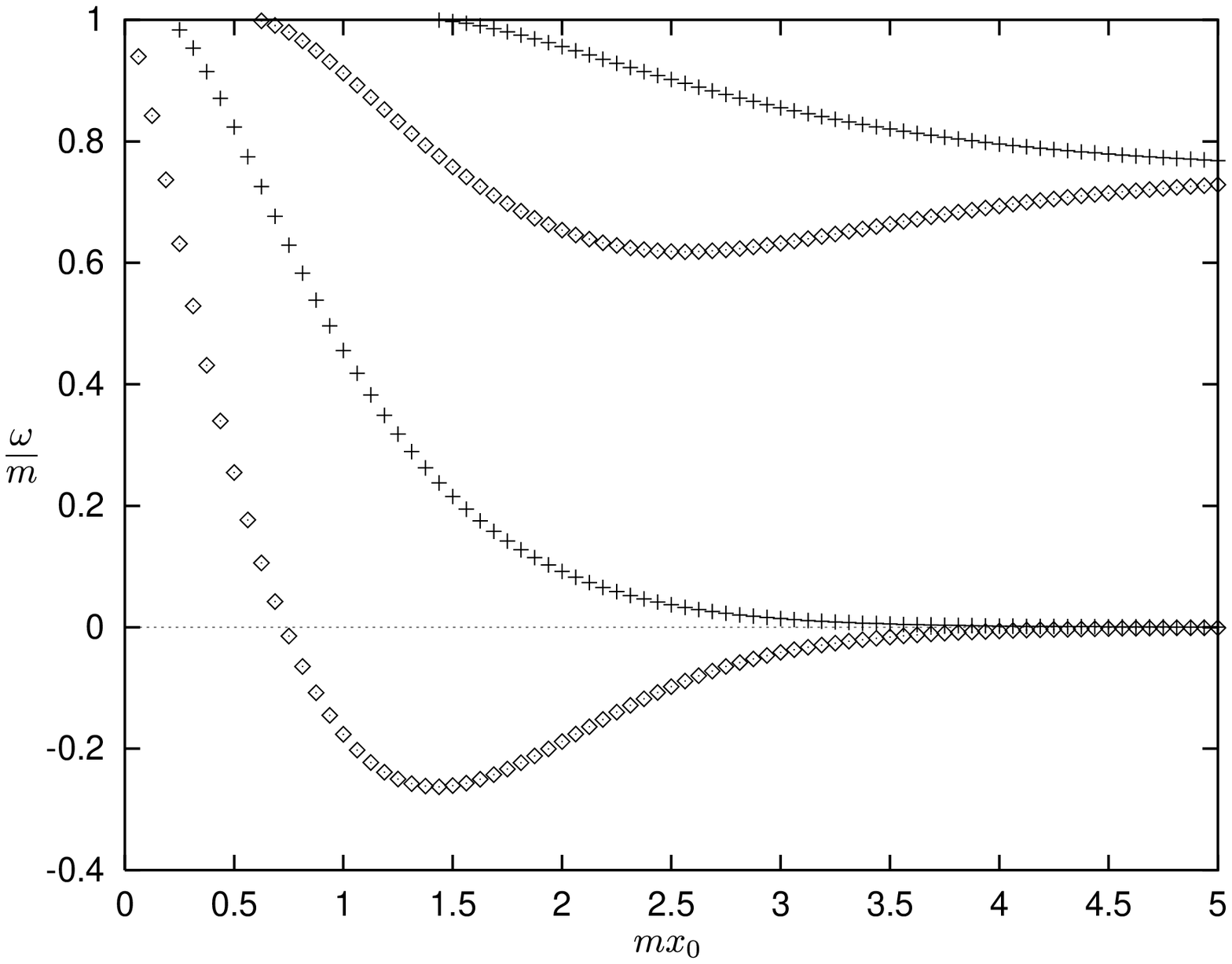,height=5.0cm,width=6.0cm} \hfill
\psfig{file=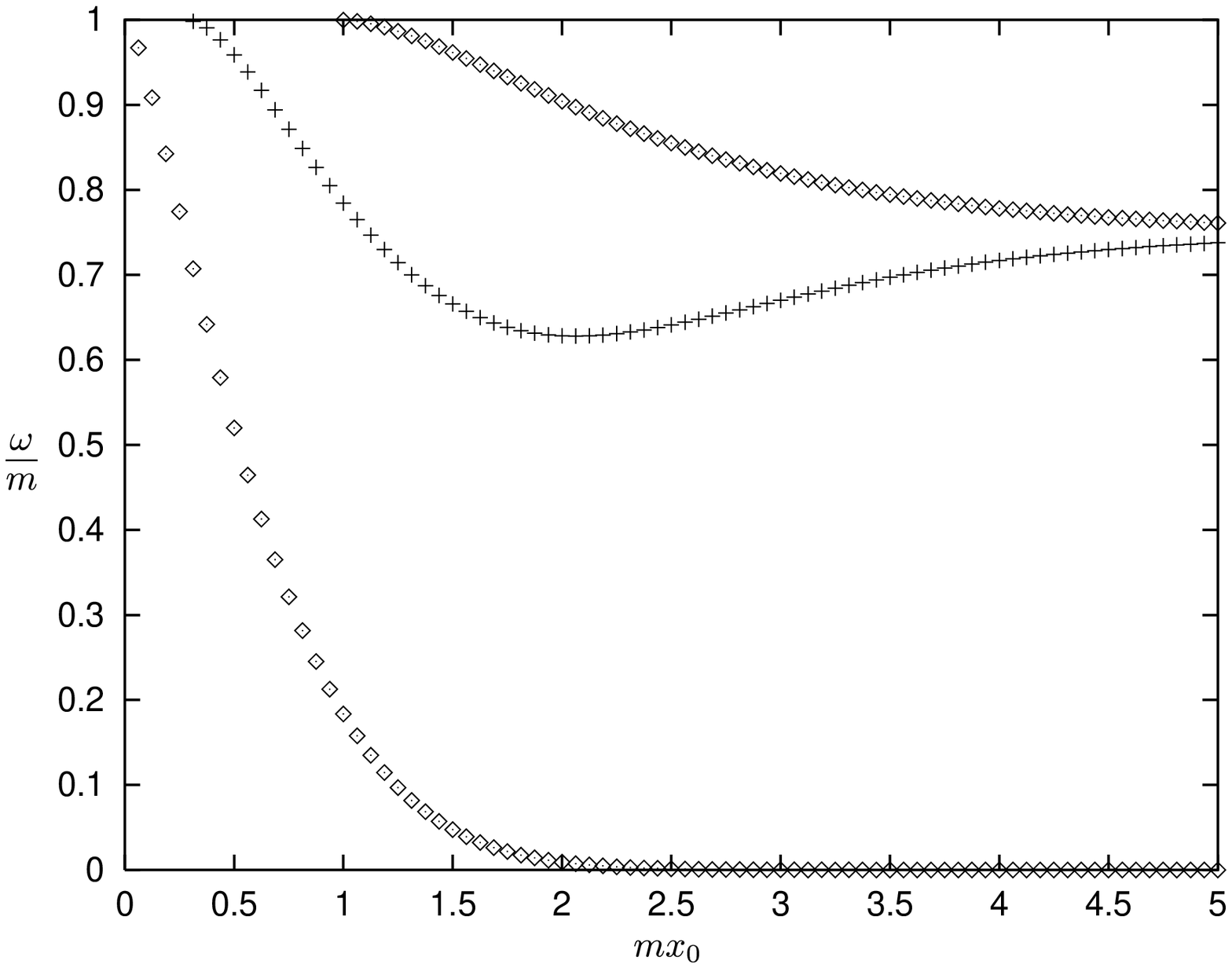,height=5.0cm,width=6.0cm}}
\fcaption{Bosonic (left) and fermionic (right) bound state spectra for
kink/antikink background with separation $2 x_0$.  We display
$\omega_B^2$ for the bosonic modes and $\omega_F$ for the fermionic
modes as functions of $x_0$. In limit of infinite separation, we have
supersymmetry, so the modes match except  for the zero modes.}
\label{spectra}
\end{figure}

To be specific, we will consider the special case of $U(\phi) =
\frac{m}{2} (\phi^2 - 1)$, where the soliton is the standard
``kink,'' $\phi_{0}(x)=\tanh mx/2$.  Our ability to consider
configurations that are not solutions to the classical equations of
motion allows us to study a sequence of background fields,
$\phi_{0}(x,x_{0})$, which interpolate between the trivial background
at $x_{0}=0$ and a widely separated kink-antikink pair as
$x_{0}\to\infty$,
\begin{equation}
    \phi_0(x,x_0) = \tanh\frac{m}{2}(x+x_0) - 
    \tanh\frac{m}{2}(x-x_0) - 1\, .
\end{equation}
This procedure enables us to avoid potential ambiguities regarding the
choice of boundary conditions when $\phi_{0}$ tends toward different
vacua as $x\to\pm\infty$ or at boundaries introduced to discretize the
problem\cite{many}.  To stabilize an arbitrary background,
$\phi_{0}(x,x_{0})$, we must insert a source term, $J(x) = \frac{d^2
\phi_0}{dx^2} - \frac{1}{2} m^2 (\phi_0^3 - \phi_0)$ into the SUSY
lagrangian, eq.~(\ref{SUSYlag}).  With this choice, $\phi_0(x,x_{0})$
is a stationary point of the action (though not necessarily a global
minimum).  The source breaks supersymmetry except when $x_{0}=0$ and
as $x_{0}\to\infty$, but it allows us to track the properties of
the system continuously from the trivial case ($x_{0}=0$) to the case
of interest ($x_{0}\to\infty$), both of which are
supersymmetric. Technical issues associated with the source,
including restoration of translation invariance and the appearance of
modes with imaginary frequencies, are discussed in
Ref.\cite{Bashinsky:1999vg}.  They do not complicate the picture
presented here.  To understand the subtleties of the renormalized
energy calculation, it is instructive to compare the bosonic and
fermionic spectra as functions of separation $x_0$ for kink-antikink
pair.  For large separation, the boson and fermion modes match, as
required by supersymmetry, except we have two boson zero modes but
only one fermion zero mode.  Figure~\ref{spectra} illustrates this
discrepancy.

Note that for each state in the
spectrum, there is another equivalent state with the opposite sign of
the energy.  Since we are considering a real scalar and a Majorana
fermion, in both cases we only consider one of these two states.

The zero modes matter even though they do not contribute directly to
the vacuum polarization energy because they are related to the
continuum through Levinson's theorem.  Since the number of bound
states is different for bosons and fermions, by Levinson's theorem the
phase shifts at $k=0$ must differ.  Since the phase shifts are
continuous, they must also differ as functions of $k$.  Indeed, for
the widely separated soliton/antisoliton pair, we find
\begin{equation}
    \delta_B(k) - \delta_F(k) = 2\arctan \frac{m}{k}
\end{equation}
and so the renormalized one--loop quantum correction to the
energy as $x_{0}\to\infty$ is
\bea
    \Delta E&=&\frac{1}{2}\sum_j\left(\omega_j^{ B }-m\right)
    -\frac{1}{2}\sum_j\left(\omega_j^{ F }-m\right)     \nonumber \\*
    && +\int_0^\infty\frac{dk}{2\pi}
    \left(\sqrt{k^2+m^2}-m\right)\frac{d}{dk}\left(\delta_B(k)-\delta_F(k)
    -\delta^{(1)}(k)\right) \cr
    &=&-\frac{m}{2} + \int_0^\infty\frac{dk}{2\pi}
    \left(\sqrt{k^2+m^2}-m\right)\frac{d}{dk}
    \left(2\arctan\frac{m}{k}-\frac{2m}{k}\right) \cr
    &=&-\frac{m}{\pi} 
    \label{susyeng}
\eea
where we have fixed the coefficient of the counterterm
\be
 {\cal L_{\rm ct}} = - C U^{\prime\prime}(\phi) U(\phi)
 - C U^{\prime\prime\prime}(\phi) \bar \Psi \Psi
\label{counterterm1}
\ee
by requiring that the tadpole graph vanish, with no further finite
renormalizations.

We assign half this energy shift to the soliton and half to the
antisoliton, so the result for the soliton is $\Delta E =
-\frac{m}{2\pi}$.  This assignment is supported by a careful
consideration of the zero modes: for a single soliton, we find one
bosonic zero mode, and ``one-half'' of a fermionic zero mode.  Just as
the bosonic zero mode reflects the breaking of translation invariance,
the fermionic mode reflects the breaking of one of the two
supersymmetry generators.  This mode is the Majorana fermion analog of
the Jackiw-Rebbi mode\cite{JR}.  For large $x_{0}$, only one mode
appears near $\omega=0$ in the spectrum of positive energy solutions
the Dirac equation.  When we reduce to the case of a single soliton,
still keeping only the positive energy states, it is weighted by one
half.  Indeed one can verify that the residue of the pole in the
fermionic Green's function at $\kappa^2 = -m^2$ is half the usual
result for a bound state.  Our result has since been confirmed using
the generalized effective action approach\cite{Dunne}.

\subsection{BPS bound}

In the supersymmetric system, there are additional restrictions on the
quantum Hamiltonian $H$.  The central charge
\be
    Z= \frac{m^2}{\lambda}\int dx\,
    U(\phi)\frac{d\phi}{dx}
    \label{defZ}
\ee
obeys the BPS bound
\be
    \langle H \rangle \ge |\langle Z \rangle |\, .
    \label{BPS}
\ee
Classically, the bound is saturated,
\be
    E_{\rm cl}=\frac{m^2}{2\lambda}\int dx\,
    \left[\left(\frac{d\phi_0}{dx}\right)^2+
    U^2(\phi_0)\right]=-\frac{m^2}{\lambda}\int dx\,
    U(\phi_0)\frac{d\phi_0}{dx}=- Z_{\rm cl}
    \label{BPScl}
\ee
and the contribution from the counterterm is equal and opposite as
well,
\be
\Delta E_{\rm ct} = C \int  U^{\prime\prime}(\phi_0) U(\phi_0)
\, dx = - C \int U^{\prime\prime}(\phi_0) \phi_0^\prime \, dx =
 - \Delta Z_{\rm ct} \,.
\ee

The negative quantum correction to the energy we found in the previous
section would appear to lead to a violation of the bound, which cannot be
correct.  The resolution is that a similar analysis, taking careful
account of renormalization by using the scattering data, yields a
compensating correction to the central charge.  Expanding the field
$\phi$ around the classical solution $\phi_0(x)$ gives
\bea
\Delta Z
&=& \langle Z \rangle_\phi - Z_{\rm cl} \cr
&=& \Delta Z_{\rm ct} + \frac{m^2}{\lambda}\int
\left\langle U' \eta \eta' - \frac{1}{2} UU''\eta^2
\right\rangle_{\phi_0} \, dx \cr
&=& \Delta Z_{\rm ct} + \frac{m^2}{2\lambda}
\int\left\langle\left((\frac{d}{dx} + U')\eta\right)^2 -
(\eta')^2 - \eta^2(U')^2 - UU''\eta^2
\right\rangle_{\phi_0} \hspace{-0.2cm} dx  \hspace{0.5cm}
\eea
where $\phi(x) = \phi_0(x) + \eta(x)$ and
\be
\eta(x) = \sqrt{\frac{\lambda}{m^2}} \left(
\int \frac{dk}{\sqrt{4\pi\omega_{k}} } \left(a_k \eta_k(x)
e^{-i\omega_k t} + a_k^\dagger \eta_k^\ast(x) e^{i\omega_k
t}\right) + \eta_{\omega=0}(x) a_{\omega=0} \right)
\ee
where $\omega_k = \sqrt{k^2 + m^2}$, the creation and annihilation
operators obey the usual commutation relations, and we have explicitly
separated the contribution of the zero mode.  The other bound states are
understood to give discrete contributions to the integral.

We can compute this expectation value and connect it to our scattering
theory formalism using the relationship between the wavefunction and the
density of states,
\be
\rho(k) - \rho_0(k) = \frac{1}{\pi} \int dx \left(|\eta_k(x)|^2 - 1\right) 
\ee
yielding as a result %
\bea
    \Delta Z&=&
    \frac{1}{4}\sum_j\left(|\tilde{\omega}_j|-m\right)
    -\frac{1}{4}\sum_j\left(|\omega_j|-m\right)
    \nonumber \\ && \hspace{1cm}
    +\int_0^\infty\frac{dk}{4\pi} \left(\sqrt{k^2+m^2}\right)
    \frac{d}{dk} \left(\tilde{\delta}(k)-\delta(k)
    +2\delta^{(1)}(k)\right)
    \nonumber \\
    &=&\frac{m}{4}-\int_0^\infty\frac{dk}{2\pi}
    \left(\sqrt{k^2+m^2}-m\right)
    \frac{d}{dk} \left(\arctan\frac{m}{k}-\frac{m}{k}\right)
    =\frac{m}{2\pi}\, ,
    \label{DZ2}
\eea
in terms of the phase shifts $\delta(k)$ and $\tilde\delta(k)$ and
the bound states energies $\omega_{j}$ and $\tilde\omega_{j}$ in the
potentials $V$ and $\tilde V$ respectively.  Comparison with
eq.~(\ref{susyeng}) shows that the correction to central charge for a
single soliton or antisoliton satisfies $|\Delta E| = |\Delta Z|$, so the
BPS bound remains saturated.  This result was subsequently confirmed by
SUSY methods\cite{SVV} to be the matrix element of an anomalous correction
to the central charge operator.

\section{Quantum Energies of Interfaces}

As a final application, we follow Ref.\cite{domainwall} and extend our
formalism to the case of interfaces, background fields that are
independent of some of the spatial coordinates, and symmetric in the
remaining ones.  Our goal will be to compute the energy per unit
transverse area, which corresponds to an induced quantum surface
tension or cosmological constant.  For an application to the case with
only one nontrivial dimension, see Ref.\cite{Yaffe} which applies
functional methods.  The general problem was considered in
Ref.\cite{Bordag}, though this approach does not make contact with
perturbative renormalization conditions.  It also requires that the
fluctuating field have an explicit mass term, so that massless fields
or fields that get their masses through spontaneous symmetry breaking
cannot be considered.

The interface is described as a background field in $n+s$ dimensions. 
The background is independent of the coordinates in $s$ ``transverse''
dimensions, but varies as a function of the $n$ ``nontrivial''
coordinates.  Examples include a domain wall in three space, where
$s=2$ and $n=1$, a magnetic vortex also in three space, where $s=1$
and $n=2$, and branes in general, where $s$ is the brane dimension and
$n$ is the codimension.  Since the background is independent of the
$s$ coordinates, the scattering phase shifts are independent of the
momentum $\vec p$ along the interface.  The energy per unit transverse
area can be written as a natural generalization of eq.~(\ref{deltaE}),
\begin{eqnarray}
        {\cal E}_{n,s}[\phi] &=&
        \pm \int \frac{d^s p}{(2\pi)^s} \sum_\ell D^\ell_n \biggl[
        \int_0^\infty \frac{dk}{2\pi}
        \left(\omega(k,p)-m(p)\right) \frac{d}{dk}
        \overline \delta_{n\ell}{}^{N}(k)
      	\nonumber \\* && \hspace{3cm}
        +\frac{1}{2}\sum_j
        \left(|\omega_{j,\ell}(p)|-m(p)\right) \biggr] +
	{\cal C}_{n,s}^N[\phi]
        \label{regularized}
\end{eqnarray}
where $m(p) = (p^2 + m^2)^{\frac{1}{2}}$,
$\omega(p,k)=(k^2+m(p)^2)^{\frac{1}{2}}$,
$\omega_j(p)=(m(p)^2-\kappa_j^2)^{\frac{1}{2}}$, and $p=|\vec p|$. 
The bound states and Born--subtracted phase shifts, $\overline
\delta_{n\ell}{}^{N}(k)$, are computed in the scattering theory
ignoring the transverse dimensions.  ${\cal C}_{n,s}^N[\phi]$
represents the Feynman diagrams corresponding to the Born
subtractions, together with the standard counterterms that renormalize
them.  Adding extra dimensions will in general require additional
subtractions beyond those associated with an $n$ dimensional
background field, since the divergences are those of a theory in $n+s$
spatial dimensions.

\begin{figure}[hbt]
\centerline{
\psfig{figure=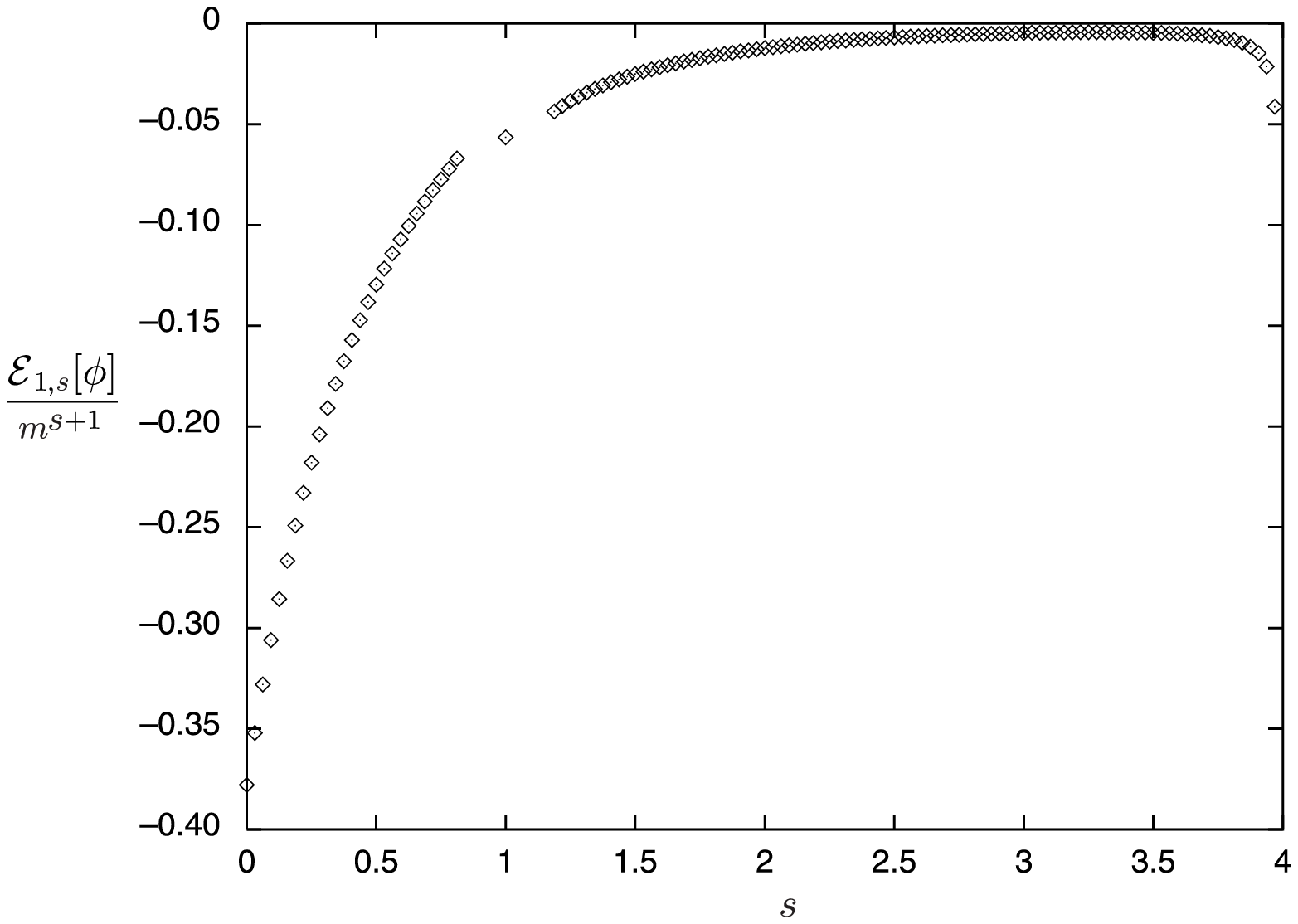,width=10cm,height=6cm}}\bigskip
\caption{\sl ${\cal E}_{1,s}[\phi]/m^{s+1}$ as a function of $s$ for a
bosonic field in the background $V(x) = -\frac{\ell+1}{\ell} m^2
{\rm sech}^2 \frac{m x}{\ell}$ with $\ell = 1.5.$
For the particular cases $s=1$ and $s=3$, the limits have been taken
analytically using the sum rule eq.~(\ref{res}) and its $s=3$
analogue\protect\cite{domainwall}.}
\label{nplot}
\end{figure}

Next we integrate over the $s$ coordinates of $p$ and obtain
\begin{eqnarray}
        {\cal E}_{n,s}[\phi] &=&
	\mp \frac{\Gamma (-\frac{s+1}{2})}{(4 \pi)^{\frac{s+1}{2}}}
        \sum_\ell D^\ell_n \biggl[
        \int_0^\infty \frac{dk}{2\pi}
        \left(\omega(k)^{s+1}-m^{s+1}\right) \frac{d}{dk}
        \overline \delta_{n\ell}{}^{N}(k)
       	\nonumber \\ && \hspace{3cm}
        +\frac{1}{2}\sum_j
        \left(|\omega_{j,\ell}|^{s+1} - m^{s+1}\right) \biggr] +
	{\cal C}_{n,s}^N[\phi]\, .
        \label{regularizedintegrated}
\end{eqnarray}
In eq.~(\ref{regularizedintegrated}), all of the divergences have been
handled with our standard prescription.  Nonetheless,
eq.~(\ref{regularizedintegrated}) appears to diverge when $s$
approaches an odd integer because of the pole in the $\Gamma$
function.  Since we know that for any smooth background all
potential divergences have been treated by Born subtraction and
subsequent renormalization, the quantity in brackets much vanish in
each channel.  Thus we are led to scattering theory sum rules, such as
\begin{equation}
    \int_0^\infty\frac{dk}{\pi}k^2\frac{d}{dk}\overline{\delta}^{(1)}(k)
    -\sum_j\kappa_j^2=0
    \label{res}
\end{equation}
which can be proved using Jost--function
techniques\cite{Puff75,sumrule}.  Using these identities, we find that
the limit where $s$ approaches an odd integer is smooth. For the case
of $s\to 1$, for example, we obtain
\begin{eqnarray}
        {\cal E}_{n,1}[\phi] &=&
	\mp \frac{1}{(4 \pi)}
        \sum_\ell D^\ell_n \biggl[
        \int_0^\infty \frac{dk}{2\pi}
        \omega(k)^2 \log \frac{\omega(k)^2}{m^2} \frac{d}{dk}
        \overline \delta_{n\ell}{}^{N}(k)
       	\nonumber \\* && \hspace{3cm}
        +\frac{1}{2}\sum_j
        \omega_{j,\ell}^2 \log \frac{\omega_{j,\ell}^2}{m^2} \biggr] +
	{\cal C}_{n,1}^N[\phi]
        \label{regularizedintegrated1}
\end{eqnarray}
which is free of divergences.  The result is independent of
the scale of the logarithms by eq.~(\ref{res}) and Levinson's theorem.
 In Figure~\ref{nplot} we consider a specific
potential for $n=1$ and study the behavior of ${\cal E}_{1,s}$
as a function of $s$.  It is smooth at $s=1$ and $s=3$, showing that
the naive divergences indeed vanish.

As a concrete example, we can apply this formula to the kink domain
wall in $2+1$ dimensions.  The Lagrangian is
\begin{equation}
    {\cal L} = \frac{m^2}{2 \lambda} \left(
    \partial_\mu \phi\partial^\mu \phi- \frac{m^2}{4}(\phi^2 - 1)^2 \right)
    + C (\phi^2 - 1)
\end{equation}
where $\phi$ is a real scalar field with mass $m$, and the counterterm
$C$ is fixed by the renormalization condition that
$\langle\phi\rangle=1$ is not renormalized, with no further finite
renormalization.  The kink solution is
\begin{equation}
    \phi_0(x) = \tanh \frac{mx}{2}
\end{equation}
and the corresponding small oscillations potential is
\begin{equation}
V(x) = -\frac{3}{2} {\rm sech}^2 mx
\end{equation}
which is an exactly solvable Posch-Teller reflectionless
potential. The total phase shift and its
first Born approximation are
\begin{eqnarray}
\delta(k) &=& 2\arctan \frac{3m}{2k} \cr
\delta^{(1)}(k) &=& \frac{3m}{k} \end{eqnarray}
and there are bound states at $\omega=0$, $\omega = m\sqrt{3}/2$ and a
``half--bound'' threshold state at $\omega = m$.  Substituting these
data into eq.~(\ref{regularizedintegrated1}) gives\cite{thermal}
\begin{equation}
{\cal E}_{1,1}[\phi_0] = \frac{3m^2}{16 \pi}\left({\rm arccoth}(2)-2\right) \,.
\end{equation}

\section{Conclusions}

We have presented a general procedure that is applicable to a variety
of problems in quantum field theory.  It gives a concrete prescription
for handling field theory divergences in a concrete way. 
Ref.\cite{qball} extends this approach to systems with simple time
dependence and Ref.~\cite{thermal} applies this formalism to nonzero
temperature.  Work is underway to apply these techniques to Higgs
solitons in the Standard Model of the weak interactions, to field
theories constrained to obey boundary conditions (such as the original
Casimir problem), and to local densities.  It could also be applied to
compute the determinants associated with instantons and bounces.

\nonumsection{Acknowledgments}
This work has been done in collaboration with E.~Farhi, P.~Haagensen,
V.~Khemani, K.~Olum, M.~Quandt, and P.~Sundberg.  N.~G. and R.~L.~J.
thank P.~van Nieuwenhuizen and A.~Vainshtein for valuable discussions
that stimulated us to carry out the calculations of Section~4.  N.~G.
is supported by the U.S.~Department of Energy (D.O.E.) under
cooperative research agreement~\#DE-FG03-91ER40662.  R.~L.~J. is
supported in part by the U.S.~Department of Energy (D.O.E.) under
cooperative research agreement~\#DF-FC02-94ER40818.  H.~W. is
supported by the Deutsche Forschungsgemeinschaft under
contracts~We~1254/3-2 and We~1254/4-2.

\bigskip
\nonumsection{References} 

\end{document}